\documentclass{jpp}
\usepackage{graphicx}
\usepackage[utf8]{inputenc}
\usepackage[T1]{fontenc}
\usepackage{amsmath,amssymb}
\usepackage{bm}
\usepackage{siunitx}
\usepackage{float}
\usepackage{relsize}
\usepackage{natbib}
\usepackage[colorlinks=true,
            linkcolor=blue,
            citecolor=blue,
            urlcolor=blue]{hyperref}
\usepackage[all]{hypcap}

\usepackage{amsmath}
\numberwithin{equation}{section}

\usepackage{orcidlink}

\shorttitle{Collisional passing alpha energy transport in nearly quasisymmetric stellarators}
\shortauthor{M. Calvo-Carrera, P.J. Catto}
\title{Collisional passing alpha energy transport in nearly quasisymmetric stellarators}
\author{Miguel Calvo-Carrera\aff{1}\orcidlink{0000-0003-3752-2657}
  \corresp{\email{miguelcc@mit.edu}},
 \and Peter J. Catto\aff{1}\orcidlink{0000-0002-0349-1736}}
\affiliation{\aff{1}Plasma Science and Fusion Center, Massachusetts Institute of Technology,
Cambridge, MA, US}

\begin{document}

\maketitle

\begin{abstract}\label{abstract}
$\!\!\!\!\!$ Recent advances in stellarator optimization have found nearly precise quasisymmetric configurations. These are expected to reduce the non-turbulent background plasma transport to acceptable neoclassical levels while removing nearly all collisionless direct orbit losses of alpha particles. Yet, alpha particles under resonant conditions can be very sensitive to collisions, causing concerning energy losses and damaging plasma facing components. For the passing alphas such resonances can happen near rational surfaces in the presence of helical error field departures from quasisymmetry that change the magnetic field direction and magnitude. The cancellation between streaming motion and tangential drift of the alphas enhances the effective collision frequency, allowing the formation of a collisional boundary layer and giving rise to a perturbed distribution function. We develop an analytical model to illustrate and evaluate the resonant plateau transport this mechanism causes by formulating a drift kinetic treatment. The results indicate the associated energy losses can become significant in the vicinity of rational surfaces at values of $q=m/n$ when error fields with poloidal and toroidal numbers $m$ and $n$ are present. In addition, we investigate the validity of the quasilinear approximation to the energy flux to show that it imposes a restriction on the error field amplitude that can be considered.
\end{abstract}
\section{Introduction}\label{introduction}

One of the critical areas for the success of fusion reactors is our ability to confine alpha particles long enough for them to transfer most of their energy to the bulk plasma. Losing too much energy from fusion born alphas would decrease the efficiency of future plants and harm the integrity of its plasma facing components. Although experiments with energetic particles from neutral beam injection (NBI) and ion cyclotron resonance heating (ICRH) have been carried out, their spatial localization and velocity space anisotropy differ from fusion born alphas (Fasoli \textit{et al.} \citeyear{fasoli_chapter_2007}; Gorelenkov \textit{et al.} \citeyear{gorelenkov_energetic_2014}). While experiments producing alpha particles have been performed, their regime is still quite far from burning plasma conditions (Villari \textit{et al.} \citeyear{villari_overview_2025}). It is therefore necessary to continue developing models to study the dynamics of alpha particles in magnetically confined fusion devices.

In recent years, quasisymmetric stellarators have become a highly promising configuration for future reactors. New optimization schemes have made it possible to find quasisymmetric configurations with error fields down to geomagnetic field levels (Landreman $\&$ Paul \citeyear{landreman_magnetic_2022}). These configurations can reduce non-turbulent transport to acceptable neoclassical levels, and have flexibility to optimize turbulent transport. At the same time, test particle guiding center collisionless and collisional alpha losses have been successfully reduced in highly optimized configurations (Landreman \textit{et al.} \citeyear{landreman_optimization_2022}). Even though prompt loss simulations sometimes take into account the effect of collisions, they do so in an initial value problem fashion, without evaluating the final steady state alpha particle distribution function due to the lack of kinetic equation details or phase space resolution to capture collisional boundary layer effects (Lazerson \textit{et al.} \citeyear{lazerson_simulating_2021}; Bader \textit{et al.} \citeyear{bader_modeling_2021}). However, alpha particles can be very sensitive to collisions under resonance conditions, causing non-negligible transport in the presence of departures from quasisymmetry (Catto \textit{et al.} \citeyear{catto_merging_2023}; Catto \citeyear{catto_collisional_2025}). In addition, when these devices are built, error fields are likely to be higher than in ideally optimized magnetic configurations.

Boundary layers appear in several contexts (Bender $\&$ Orszag \citeyear{bender_advanced_1999}). In drift kinetic transport, collisional boundary layers are relevant in neoclassical transport (Calto \textit{et al.} \citeyear{calvo_effect_2017}) and trapped alpha particle resonant transport due to magnetic field errors in tokamaks (Catto \citeyear{catto_collisional_2019}) and quasisymmetric stellarators (Catto \textit{et al.} \citeyear{catto_merging_2023}). Tolman $\&$ Catto (\citeyear{tolman_drift_2021}) also show possible trapped particle transport in the presence of time dependent perturbations with an effective Krook collision operator for an axisymmetric configuration. In this work, we evaluate the resonant energy transport of passing alpha particles in the presence of steady state deviations from quasisymmetry in magnitude and field direction. To resolve the collisional boundary layer that forms around resonances with these perturbations, we use the diffusive pitch angle scattering term due to alpha collisions with the background ions from the collision operator.

In particular, we first formulate a drift kinetic treatment of alpha particles. For passing alphas, the resonance between the parallel streaming and the tangential drift terms leads to the formation of a collisional boundary layer where the effective collisionality is enhanced and a well behaved perturbed distribution function exists. These particles are sensitive to collisions and can cause substantial radial transport. By working in an appropriate set of coordinates and trajectory averaging along the streaming variable, we arrive at a 1D equation in pitch angle. This procedure allows us to calculate the perturbed distribution function, which we input into the quasilinear flux expression obtained from the lowest order drift kinetic equation to get the radial energy flux and diffusivity. The alpha energy diffusivity allows us to assess potential energy losses and estimate the limitations of the model. 

In section \ref{section2} we describe the magnetic field background and spatial coordinates, and the deviation from quasisymmetry. In section \ref{section3} the model is introduced, including the lowest and first order equations, the solution to the perturbed distribution function and its resulting energy transport. In section \ref{section4} phenomenological estimates are included, which provide physical intuition into collisional boundary layers, and an explanation of the limitations of the model is presented. Section \ref{section5} shows the evaluation of the resulting transport, along with a study of the model limitations and a sensitivity analysis of the key parameters. In section \ref{discussion} the results are discussed, including some insights for stellarator design and conclusions. In the Appendices, we discuss the barely passing asymptotic limit and the agreement with previous results, provide some additional mathematical details, and include a more detailed discussion on the model limitations.

\section{Magnetic field description}\label{section2}
    \subsection{Background magnetic field}\label{subsection21}

A general stellarator magnetic field with well defined nested flux surfaces can be described by (Boozer \citeyear{boozer_plasma_1981})
\begin{equation}
\mathbf{B} = B\,\mathbf{b}
= \nabla \alpha \times \nabla \psi_p
= K\,\nabla \psi_p + G\,\nabla \vartheta + I\,\nabla \zeta .
\label{2.1}
\end{equation}
where $\vartheta$ is the poloidal angle and $\zeta$ is the toroidal angle. The poloidal flux function $\psi_p$, defined by $2\pi \psi_p = \int_{\mathrm{pol}} d^{2}\bm{r}\cdot\bm{B}$, labels the magnetic surfaces, and along with $\alpha \equiv \zeta - q(\psi_p)\,\vartheta$, defines the magnetic field lines in every flux surface. The safety factor $q ={\bm{B}\cdot\nabla \zeta}/{\bm{B}\cdot\nabla \vartheta}$
 has been defined to be a flux function, and the rotational transform is $t = q^{-1}$. The coefficients $G$ and $I$ are defined as the toroidal and poloidal currents, as in Landreman $\&$ Catto (\citeyear{landreman_omnigenity_2012})
\begin{equation}
\int_{\mathrm{tor}} d^{2}\bm{r}\cdot\bm{J}
= \frac{cG}{4\pi}\int_{-\pi}^{\pi} d\bm{r}\cdot\nabla \vartheta
= \frac{cG}{2},
\label{2.2}
\end{equation}
\begin{equation}
\int_{\mathrm{pol}} d^{2}\bm{r}\cdot\bm{J}
= \frac{cI}{4\pi}\int_{0}^{2\pi} d\bm{r}\cdot\nabla \zeta
= \frac{cI}{2},
\label{2.3}
\end{equation}
where $\bm{J}$ is the current and $c$ is the speed of light. The definition of $\bm{B}$ gives $\bm{B} \cdot \nabla  \psi_{p} = 0 = \bm{B} \cdot \nabla  \alpha$. Forming the product of the second and third forms of $\bm{B}$ in \eqref{2.1} we calculate the inverse of the Jacobian of the set of coordinates $(\psi_{p},\vartheta,\zeta)$
\begin{equation}
\nabla \psi_p \times \nabla \vartheta \,\cdot\, \nabla \zeta
= \bm{B}\cdot\nabla \vartheta
= q^{-1}\,\bm{B}\cdot\nabla \zeta
= {B^{2}}/{(q I + G)}.
\label{2.4}
\end{equation}
From pressure balance $\bm{J}\times\bm{B}=c\nabla  p(\psi_p)$, where $p$ is the pressure, the radial current $\bm{J}\cdot\nabla \psi_p$ has to vanish, giving $\nabla \cdot(\bm{B}\times\nabla \psi_p)=0$. This condition forces the coefficients in $\bm{B}$ to satisfy $\partial I/\partial\vartheta=\partial G/\partial\zeta$ and therefore
${\partial}(\oint d\zeta\,I)/{\partial\vartheta} = 0 =
{\partial}(\oint d\vartheta\,G)/{\partial\zeta}$.
Consequently, this suggests working in Boozer coordinates with $I=I(\psi_p)$ and $G=G(\psi_p)$ flux functions (Boozer \citeyear{boozer_plasma_1981}) and the Jacobian’s angular dependence that of $B^2$. Hence, the Lagrangian of the motion of charged particles in this field will inherit the angular dependence of $B^2$. In an ideal quasisymmetric stellarator field there is an angular coordinate that $B=|\vec{B}|$ does not depend on, implying by Noether’s theorem (Noether \citeyear{noether_invariante_1918}) that we will be able to define a conserved canonical momentum. Satisfying this property, as in axisymmetric fields, ensures the collisionless confinement of particles in a more general class of configurations, known as quasisymmetric (QS) stellarators.

    \subsection{Spatial coordinates}\label{subsection22}

We work in a quasisymmetric (QS) configuration with quasisymmetric poloidal and toroidal numbers $M$ and $N$ positive integers. The magnetic field magnitude on a flux surface only depends on a single periodic angle variable (Nührenberg $\&$ Zille \citeyear{nuhrenberg_quasi-helically_1988}; Boozer \citeyear{boozer_quasi-helical_1995})
\begin{equation}
  \eta = M\vartheta - N\zeta.
\label{2.5}
\end{equation}
To take advantage of the symmetries of the problem, it is often convenient to work with the angular variables $\eta,\alpha$ instead of the Boozer angles $\vartheta,\zeta$. They relate to each other as
\begin{equation}
    \vartheta = {(N\alpha + \eta)}/{(M - qN)},
\label{2.6}
\end{equation}
\[\\[-22pt]\]
\begin{equation}
    \zeta = {(M\alpha + q\eta)}/{(M - qN)}.
\label{2.7}
\end{equation}
In the set of coordinates $(\psi_p,\eta,\alpha)$ the inverse of the Jacobian becomes
\begin{equation}
    \nabla \psi_p \times \nabla \eta \cdot \nabla \alpha
    = \bm{B}\cdot\nabla \eta
    = (M - qN)\,\bm{B}\cdot\nabla \vartheta
    = (M - qN)\,B^{2}/(q I + G)
\label{2.8}
\end{equation}

In the radial direction, we define the toroidal flux function $\psi_t$ as $2\pi \psi_t = \int_{\mathrm{tor}} d^{2}\bm{r}\cdot\bm{B}$, which satisfies $\partial \psi_{t} / \partial \psi_{p} = q(\psi_{p})$, and the helical flux function as
\begin{equation}
    \psi_{h} = M\psi_{p}-N\psi_{t}.
\label{2.9}
\end{equation}
Differentiating \eqref{2.9} we get $d\psi_h = (M - qN)\, d\psi_p$, and therefore the inverse of the Jacobian of the set of coordinates $\nabla \psi_h \times \nabla \eta \cdot \nabla \alpha = (M - qN)\,\bm{B}\cdot\nabla \eta$. At times it will be useful to work with the drift kinetic canonical helical momentum as it is an approximate constant of the motion (Boozer \citeyear{boozer_transport_1983}; Landreman $\&$ Catto \citeyear{landreman_effects_2010})
\begin{equation}
    \psi_* = \psi_h - I_h\,{v_{\parallel}}/{\Omega}.
\label{2.10}
\end{equation}
where $I_h \equiv MI + NG$. The gyrofrequency is $\Omega = Z_{\alpha}eB/M_\alpha c$, with $Z_{\alpha}$ and $M_\alpha$ the alpha charge number and mass, and $e$ the proton charge. The parallel velocity $v_{\parallel}$ is obtained from the alpha particle energy $E = {v^{2}}/{2} + {Z_{\alpha} e\,\Phi}/{M_{\alpha}}$, to get $v_{\parallel}^{2} = 2( E - {Z_{\alpha} e\,\Phi}/{M_{\alpha}} - \mu B)$, with $\Phi$ the electrostatic potential, $\mu = {v_{\perp}^{2}}/{2B}$ the alpha particle magnetic moment and $v_\perp$ its perpendicular velocity. As shown in Appendix~\ref{appendixA}, in an ideal QS magnetic field $\psi_*$ satisfies
\begin{equation}
    \big( v_{\parallel}\,\bm{b} + \bm{v}_d \big)\cdot\nabla \psi_* = 0,
\label{2.11}
\end{equation}
and to the requisite order the drift velocity can be taken to be
\begin{equation}
\bm{v}_d
= \frac{1}{\Omega}\,\bm{b}\times\!\big( \mu\,\nabla  B
  + v_{\parallel}^{2}\,\bm{b}\cdot\!\nabla \bm{b} \big)
  + \frac{c}{B}\,\bm{b}\times\nabla \Phi
\approx
\frac{v_{\parallel}}{\Omega}\,
(\bm{\mathsf{I}}-\bm{b}\bm{b})\cdot\!\nabla \times
(v_{\parallel}\bm{b})
\approx
\frac{v_{\parallel}}{\Omega}\,
\nabla \times(v_{\parallel}\bm{b}) .
\label{2.12}
\end{equation}
with $\bm{\mathsf{I}}$ the unit dyad. In \eqref{2.11}, the parallel velocity correction $u_{\parallel} = -\,\Omega^{-1}\, v_{\perp}^{2}\,\bm{b}\cdot\nabla \times\bm{b}$ (Hazeltine \citeyear{hazeltine_recursive_1973}; Catto \citeyear{catto_linearized_1978}) is neglected as small compared to $v_{\parallel}$, as discussed in  Boozer (\citeyear{boozer_guiding_1980}); Parra $\&$ Catto (\citeyear{parra_limitations_2008}); Landreman $\&$ Catto (\citeyear{landreman_conservation_2013}).
  
    \subsection{The magnetic perturbation}\label{subsection23}

To evaluate the effect of error fields on the alpha distribution function we introduce subscripts to distinguish between the quasisymmetric (QS) magnetic field $\bm{B}_0 = B_0(\psi_p,\eta)\,\bm{b}_0$ and a general helical perturbed magnetic field $\bm{B} = \bm{B}_0 + \bm{B}_1$ with
\begin{equation}
\bm{B}_1 = \nabla \times\bm{A}_1
= B_{\parallel}\,\bm{b}_0
+ \big( \bm{\mathsf{I}} - \bm{b}_0\bm{b}_0 \big)
   \cdot\nabla \times
   \big( A_{\parallel}\,\bm{b}_0
       + A_{\alpha}\,\nabla \alpha
       + A_{\psi}\,\nabla \psi_p \big) .
\end{equation}
Here $\bm{A}_1 = A_{\parallel}\,\bm{b}_0 + A_{\alpha}\,\nabla \alpha + A_{\psi}\,\nabla \psi_p$ is the perturbed vector potential, and $B_{\parallel} = \bm{b}_0 \cdot \nabla \times\bm{A}_1$ is the departure from quasisymmetry of the magnetic field magnitude, with $B=B_0 + B_{\parallel}$. The unit vector $\bm{b}_0$ is deflected by $\bm{B}_1/B_0$ due to $\bm{A}_1$, with $\psi_p$ being perturbed by $\psi_1$. If the result were perfect flux surfaces, then $\bm{B}\cdot\nabla \psi \to 0$, and since $\bm{B}_0\cdot\nabla \psi_p = 0$, this would require  $\bm{B}_0\cdot\nabla \psi_1 + \bm{B}_1\cdot\nabla \psi_p \to 0$. However, forming $\bm{B}_1\cdot\nabla \psi_p$ gives
\begin{equation}
\bm{B}_1\cdot\nabla \psi_p
= \nabla \cdot\left( \frac{A_{\parallel}}{B_0}\, \bm{B}_0\!\times\!\nabla \psi_p \right)
  \!+\! \bm{B}_0\cdot\nabla  A_{\alpha}
\!=\! -\! \,B_0^{2}\,\frac{\partial}{\partial\alpha}\!\left( \frac{A_{\parallel}}{B_0} \right)
  \!+\! \bm{B}_0 \cdot\nabla \left[
      A_{\alpha}
      \!+\! \frac{I_h}{M \!-\! qN}\,\frac{A_{\parallel}}{B_0}
    \right] .
\label{2.14}
\end{equation}
where we have used $\bm{B}_0\times\nabla \psi_p\cdot\nabla \eta
  = [{I_h}/({M - qN})]\,\bm{B}_0\cdot\nabla \eta$.
In the absence of an $\alpha$ dependence of the magnetic field, we should recover
$\bm{B}_1\cdot\nabla \psi_p + \bm{B}_0\cdot\nabla \psi_1 = 0$, so we must pick $\psi_1 =-\{ { A_{\alpha}
  + [{I_h}/({M - qN})]\,{A_{\parallel}}/{B_0} } \}$ 
to obtain
\begin{equation}
\bm{B}_1\cdot\nabla \psi_p + \bm{B}_0\cdot\nabla \psi_1
= -\,B_0\,{\partial A_{\parallel}}/{\partial\alpha} .
\label{2.15}
\end{equation}
The RHS causes a small ripple in the flux surface. However, as discussed in section \ref{section4}, collisions prevent this ripple from trapping particles. This departure from perfect flux surfaces is well behaved since a periodic perturbation gives
\begin{equation}
\langle \bm{B}_1\cdot\nabla \psi_p + \bm{B}_0\cdot\nabla \psi_1 \rangle
= \langle B_0\,\partial A_{\parallel} / \partial\alpha \rangle = 0 .
\label{2.16}
\end{equation}
where the background field has been used in the flux-surface average, defined as
\begin{equation}
\langle\cdots\rangle
\equiv  {\oint \frac{d\vartheta\,d\zeta\,(\cdots)}{\bm{B}\cdot\nabla \vartheta}} \Big/ {\oint \frac{d\vartheta\,d\zeta\,}{\bm{B}\cdot\nabla \vartheta}}
= {\oint \frac{d\eta\,d\alpha\,(\cdots)}{\bm{B}\cdot\nabla \eta}} \Big/ {\oint \frac{d\eta\,d\alpha\,}{\bm{B}\cdot\nabla \eta}}.
\label{2.17}
\end{equation}

The radial departure from quasisymmetry can be estimated as
\begin{equation}
B_{\perp} \equiv 
{ \left| B_0\,\partial A_{\parallel}/\partial\alpha \right| }/
     { |\nabla \psi_p| } ,
\label{2.18}
\end{equation}
where $2\pi\psi_t \approx \bar{B}_0\,a$, with 
$a \equiv \pi r^{2}$ the area inside the elliptical flux surface about the magnetic axis
labeled by $r$, and $a' = \partial a / \partial r$. As a result,
$|\nabla \psi_p| \approx \bar{B}_0\,a'/2\pi q$.
The magnetic perturbation induces a departure from the quasisymmetric background,
$B_0 = B_0(\psi_p,\eta)$, causing $\alpha$ variation in both the streaming and drift
terms of the drift kinetic equation. This produces a perturbed distribution function
and drives radial transport.

\section{Resonant plateau transport}\label{section3}
    \subsection{Quasilinear treatment of the collisional radial transport of alpha particles}\label{subsection31}
The drift kinetic equation for alpha particles in
$\bm{r}$, $E$, and $\mu$ variables is
\begin{equation}
{\partial f}/{\partial t}
+ (v_{\parallel}\bm{b} + \bm{v}_d)\cdot\nabla  f
= C\{f\} + S\,\delta(v - v_0)/4\pi v^{2},
\label{3.1}
\end{equation}
with an alpha birth rate 
$S = n_D n_T \langle\sigma v\rangle_{DT}$,
$v_0$ the alpha birth speed, 
$n_D$ and $n_T$ the deuterium and tritium densities, and
$\langle\sigma v\rangle_{DT}$ their fusion reaction rate.
Subscripts for species are $e$ for electrons, 
$\alpha$ for alpha particles, and $i$ for the background ions that are deuterium
$D$ and tritium $T$. The alpha collision operator $C\{f\}$ is obtained by expanding the unlike particle
operator for $v_e \gg v_0 \gg v_i$, with 
$v_e = (2T_e/m_e)^{1/2}$
and $v_i = (2T_i/M_i)^{1/2}$ the electron and ion thermal speeds, $T_e$ and $T_i$ their temperatures and $m_e$ and $M_i$ their masses.  
The result is
\begin{equation}
C\{f\}
= \tau_s^{-1}\,\nabla _v\cdot\!\left[
  (v^{3} + v_c^{3})v^{-3}\bm{v}\,f
  + ({v_{\lambda}^{3}}/{2v^{3}})
    \left( v^{2}\bm{\mathsf{I}} - \bm{v}\bm{v} \right)\cdot\nabla _v f
\right],
\label{3.2}
\end{equation}
with $\tau_s ={3 M_{\alpha} T_e^{3/2}}/
               {4 (2\pi m_e)^{1/2} Z_{\alpha}^{2} e^{4} n_e \ln\Lambda}$
the slowing down time for alpha particles to thermalize to He ash and $\ln\Lambda$ the Coulomb logarithm. The first piece of the collision operator slows down the distribution function, with electron drag dominating over ion drag of birth alphas for $v>v_c$ and $v_c$ the critical speed for ion drag to equal electron drag, where $v_c^3=[3\pi^{1/2} T_e^{3/2}/(2m_e)^{1/2} n_e]\sum_i Z_i^2 n_i/M_i$. The last term in \eqref{3.2} makes the distribution function isotropic in velocity space through pitch angle scattering by ions, with a multiplying factor $v_{\lambda}^3=[3\pi^{1/2} T_e^{3/2}/(2m_e)^{1/2} M_{\alpha} n_e]\sum_i Z_i^2 n_i \sim v_c^3$.

We assume a quasisymmetric background $B_0=B_0(\psi_p,\eta)$ with a small error field that introduces an $\alpha$ dependent steady state perturbation. We demonstrate that this perturbation will create a pitch angle dependent perturbed distribution function $f_1$ that causes transport. In the $f_0$ equation, to lowest order it is convenient to allow the streaming, collision and source terms compete. Noting that $v_{\parallel} \bm{b} \cdot \nabla (\dots)=v_{\parallel} \bm{b} \cdot \nabla \eta\, \partial(\dots)/\partial\eta$, we define the transit average as
\begin{equation}
\langle \dots \rangle_{\tau} \equiv \frac{\oint d\tau\, ( \dots) }{\oint d\tau}
= \oint \frac{d\eta ( \dots)}{v_{\parallel} \bm{b} \cdot \nabla \eta} \Big/ 
      \oint \frac{d\eta}{v_{\parallel} \bm{b} \cdot \nabla \eta},
\label{3.3}
\end{equation}
where $d\tau=d\eta/v_{\parallel} \bm{b} \cdot \nabla \eta$ and $\alpha$ is kept constant. Transit averaging the steady state lowest order equation of \eqref{3.1} eliminates the streaming term, giving the solubility constraint $\langle C\{f_0\}+S\delta(v-v_0)/4\pi v^2 \rangle_{\tau}=0$, which has as solution the slowing down tail distribution
\begin{equation}
f_0 = f_0(\psi_p,E) = S\tau_s H(v_0-v)/4\pi(v^3+v_c^3),
\label{3.4}
\end{equation}
where $H(v_0-v)$ is the Heaviside step function that vanishes for $v>v_0$. Since \eqref{3.4} satisfies $C\{f_0\}+S\delta(v-v_0)/4\pi v^2=0$, then $v_{\parallel} \bm{b}_0 \cdot \nabla  f_0 = v_{\parallel} \bm{b}_0 \cdot \nabla \eta\, \partial f_0/\partial\eta = 0$, confirming that in the presence of radial losses, the lowest order distribution $f_0$ depends at most on $\psi_p$ and $E$. The smallness of the perturbation allows us to obtain $f_1$ from a linearized kinetic equation. The range of validity of this framework is evaluated a posteriori. 

Using the last form of \eqref{2.12}, the drift kinetic equation can be expressed as
\begin{equation}
C\{f\}+\frac{S\delta(v-v_0)}{4\pi v^2} = (v_{\parallel} \bm{b} + \bm{v_d}) \cdot \nabla  f 
= \frac{v_{\parallel}}{B} \, \nabla  \cdot \Big[ f(\bm{B} + \frac{B}{v_{\parallel}}\, \bm{v_d}) \Big].
\label{3.5}
\end{equation}
We transit average \eqref{3.5} and average over $\alpha$ using $\langle \dots \rangle_{\alpha} \equiv \oint d\alpha\,(\dots)/\oint d\alpha$. Linearizing $f=f_0+f_1$, with $\langle f_1 \rangle_{\alpha} = 0$ to the requisite order, we obtain
\begin{equation}
\oint d\tau\, C\{f_0\} 
+ \frac{S\delta(v-v_0)}{4\pi v^2} \oint d\tau
= \Big\langle \oint d\tau\, \frac{v_{\parallel}}{B}\, \nabla \cdot\Big[ f(\bm{B} + \frac{B}{v_{\parallel}}\, \bm{v_d}) \Big] \Big\rangle_{\alpha},
\label{3.6}
\end{equation}
where the $f_0$ piece on the right side will be proven to vanish. To obtain an expression for the right side, we neglect the neoclassical drive term, since it has been considered in previous studies (Landreman $\&$ Catto \citeyear{landreman_effects_2010}), and we focus on transport caused by departures from quasisymmetry with $\alpha$ dependence. The perturbation of $\bm{B}=\bm{B}_0+\bm{B}_1$ and $\psi=\psi_p+\psi_1$ changes the streaming term, which becomes $v_{\parallel} (\bm{b}_0 + B_0^{-1} \bm{B}_1)\cdot\nabla \psi_p + v_{\parallel} \bm{b}_0\cdot\nabla \psi_1$. This modifies the quasilinear equation through
\begin{equation}
(\bm{B} + \frac{B}{v_{\parallel}} \bm{v}_d) \cdot \nabla \psi 
\approx \bm{B}_0 \cdot \nabla \psi_1 + \bm{B}_1 \cdot \nabla \psi_p 
+ \frac{B_0}{v_{\parallel}}\, \bm{v}_d \cdot \nabla \psi_p
= -B_0\, \frac{\partial A_{\parallel}}{\partial\alpha}
  - B_0^2\, \frac{\partial}{\partial\alpha} \Big( \frac{v_{\parallel}}{\Omega} \Big).
\label{3.7}
\end{equation}
The first term in the final form captures the effect on the streaming term of the perturbation in the field direction and the flux function as in \eqref{2.15}. The second includes the effect on the drift from the departure of the magnetic field magnitude from QS, and is given in \eqref{A2} along with the ignored neoclassical term. Using the definition of $v_{\parallel}$ and neglecting the effect of the electrostatic potential $\Phi$ on the alpha particles since $e\Phi \sim T_i$, we express the second term as
\begin{equation}
v_{\parallel} B_0\, \frac{\partial}{\partial\alpha} \Big( \frac{v_{\parallel}}{\Omega} \Big) = -\frac{\mu_0 B_0 + v_{\parallel}^2}{B_0} \, \frac{\partial B}{\partial\alpha}
= -\frac{\mu_0 B_0 + v_{\parallel}^2}{B_0} \, \frac{\partial B_{\parallel}}{\partial\alpha}.
\label{3.8}
\end{equation}
Notice both terms on the right side of \eqref{3.7} then vanish when flux surface averaged. Using the property $\langle \nabla \cdot(\dots) \rangle = ({1}/{V'} )\partial [ V' \langle (\dots)\cdot\nabla \psi \rangle ]/\partial\psi$ with $V' = \oint d\eta d\alpha/\bm{B} \cdot \nabla \eta$, makes the $f_0$ term on the RHS of \eqref{3.6} also vanish in the presence of sufficiently well defined flux surfaces
\begin{equation}
\oint \frac{ d\eta\, d\alpha }{\bm{B}_0 \cdot \nabla \eta} 
\nabla \cdot \Big[f_0 (\bm{B} + \frac{B}{v_{\parallel}} \bm{v}_d) \Big]
= \frac{\partial}{\partial\psi} \Big[ f_0 \oint \frac{d\eta d\alpha}{\bm{B}_0 \cdot \nabla \eta} (\bm{B} + \frac{B}{v_{\parallel}} \bm{v}_d)\cdot\nabla \psi \Big] = 0.
\label{3.9}
\end{equation}
Therefore \eqref{3.6}  becomes
\begin{equation}
\oint d\tau\, C\{f_0\}
+ \frac{S\delta(v-v_0)}{4\pi v^2}\oint d\tau
= -\frac{\partial}{\partial\psi_p}
 \Big\{ \oint d\tau\, v_{\parallel} 
\Big\langle f_1 \Big[ \frac{\partial A_{\parallel}}{\partial\alpha}
 + B_0\, \frac{\partial}{\partial\alpha} \Big( \frac{\partial v_{\parallel}}{\partial \Omega_0} \Big) \Big] \Big\rangle_{\alpha} \Big\},
\label{3.10}
\end{equation}
where for a quasisymmetric configuration 
$V' = (qI+G)(M-qN)^{-1} \oint d\eta d\alpha/B_0^2 \approx 4\pi^2 qR/B_0$, 
since $\oint d\zeta d\vartheta=(2\pi)^2=(M-qN)^{-1} \oint d\eta d\alpha$, and we assume $I \approx \bar{B}_0 R$ with a large toroidal field $qI\gg G$. By calculating $f_1$ and integrating over velocity space, we will estimate the energy losses.
   
    \subsection{The $f_1$ equation and collisional boundary layers}\label{subsection32}

Following similar steps as in subsection \ref{subsection31}, the linearized equation for $f_1$ is
\begin{equation}
( v_{\parallel} \bm{b}_0 + \bm{v}_d ) \cdot \nabla  f_1 - C\{ f_1 \}
= \left[ v_{\parallel} \frac{\partial A_{\parallel}}{\partial \alpha}
+ v_{\parallel} B \frac{\partial ( v_{\parallel}/\Omega )}{\partial \alpha} \right]
\frac{\partial f_0}{\partial \psi_p}.
\label{3.11}
\end{equation}
Note the quasilinear treatment permits $\partial f_1 / \partial \psi_p$ to drive transport on the right side of the $f_0$ equation \eqref{3.10} by neglecting the nonlinear term in the linearized $f_1$ equation \eqref{3.11}. The limitations of this assumption are discussed in section \ref{section4}. We solve \eqref{3.11} by employing unperturbed QS trajectories on the left side. Using the angular variable $\alpha_* = \zeta - q(\psi_*) \vartheta$ to avoid a secular behavior in the tangential drift (Catto \textit{et al.} \citeyear{catto_merging_2023}), and $\psi_*$ as it is a constant of the motion in the QS field, the equation for $f_1(\psi_*,\eta,\alpha_*,E,\mu,\sigma)$ becomes
\begin{equation}
( v_{\parallel} \bm{b}_0 + \bm{v}_d ) \cdot
\Big( \nabla \eta \, \frac{\partial f_1}{\partial\eta}
+ \nabla \alpha_* \, \frac{\partial f_1}{\partial\alpha_*} \Big)
- C\{ f_1 \}
=
\left[
v_{\parallel} \frac{\partial A_{\parallel}}{\partial\alpha}
+ v_{\parallel} B \frac{\partial}{\partial\alpha} \Big( \frac{v_{\parallel}}{\Omega} \Big)
\right]
\frac{\partial f_0}{\partial \psi_p}.
\label{3.12}
\end{equation}
The left side terms are calculated in Appendix~\ref{appendixA}. The tangential drift across field lines is
\begin{equation}
\begin{aligned}
\omega_{\alpha} &\equiv ( v_{\parallel} \bm{b}_0 + \bm{v}_d ) \cdot \nabla \alpha_*
\\
&\approx
v_{\parallel} \bm{b}_0 \cdot \nabla \eta
\left[
q_* \frac{\partial}{\partial\psi_h} \left( \frac{I v_{\parallel}}{\Omega_0} \right)
+ \frac{\partial}{\partial\psi_h} \left( \frac{G v_{\parallel}}{\Omega_0} \right)
- \frac{q_* - q}{M - qN}
\right]
- v_{\parallel} \bm{b}_0 \cdot \nabla \Big( \frac{K_p v_{\parallel}}{\Omega_0} \Big),
\end{aligned}
\label{3.13}
\end{equation}
while the motion along the field lines is
\begin{equation}
( v_{\parallel} \bm{b}_0 + \bm{v}_d ) \cdot \nabla \eta
= v_{\parallel} \bm{b}_0 \cdot \nabla \eta
\left[ 1 - \frac{\partial}{\partial\psi_h} \Big( \frac{I_h v_{\parallel}}{\Omega} \Big) \right]
\approx v_{\parallel} \bm{b}_0 \cdot \nabla \eta.
\label{3.14}
\end{equation}
The second term on the right side of \eqref{3.14} can be neglected under the assumption
$\partial( I_h v_{\parallel}/\Omega ) / \partial\psi_h \sim q v/\Omega_0 a' \ll 1$, with
$R\, \partial/\partial r \sim 1$ and $R$ the nominal major radius of the magnetic axis. The equation in $\psi_*,\eta,\alpha_*,E$ and $\mu$ variables then becomes
\begin{equation}
v_{\parallel} \bm{b} \cdot \nabla \eta\, \frac{\partial f_1}{\partial\eta}
+ \omega_{\alpha} \frac{\partial f_1}{\partial\alpha_*}
- C\{ f_1 \}
=
v_{\parallel}
\left[
\frac{\partial A_{\parallel}}{\partial\alpha}
+ B_0 \frac{\partial}{\partial\alpha} \Big( \frac{ v_{\parallel}}{\Omega} \Big)
\right]
\frac{\partial f_0}{\partial\psi_p}.
\label{3.15}
\end{equation}

Since we are interested in evaluating diffusive alpha energy loss before alpha particles are slowed down by the bulk plasma, the relevant timescale will be faster than the slowing down time. Therefore, in \eqref{3.15} we can neglect the drag piece of the collision operator. The losses of concern happen when a resonance between the streaming and drift terms occurs. Since the drift depends on $v_{\parallel}$ and $v_{\perp}$, and it has to compete with the streaming, the most sensitive parameter in the resonance condition will be the pitch angle, defined as $\lambda = 2\mu_0 \bar{B}_0/v^2 = v_{\perp}^2 \bar{B}_0 /  v^2 B_0 $, which enters in $v_{\parallel}^2 = v^2 (1 - \lambda B_0/\bar{B}_0)$, with $\bar{B}_0^2 = \langle B^2 \rangle$. In this variable, our collision operator becomes
\begin{equation}
C_{\mathrm{pas}}\{ f_1 \}
=
\frac{v_{\lambda}^3 }{ 2 \tau_s }
\nabla _v \cdot [ v^{-3} ( v^2 \bm{\mathsf{I}} - \bm{v}\bm{v} ) \cdot \nabla _v f_1 ]
=
\frac{ 2 v_{\lambda}^3 \bar{B}_0 v_{\parallel} }{ \tau_s B_0 v^4 }
\frac{\partial}{\partial\lambda}
\left( \frac{\lambda v_{\parallel}}{v} \, \frac{\partial f_1}{\partial\lambda} \right).
\label{3.16}
\end{equation}
For most alphas, streaming balances the drive on the right side of \eqref{3.15}, keeping $f_1$ small and collisions unimportant, since alpha particles are born at very high speeds. However, when a resonance occurs the streaming plus drift term vanishes, leaving only the collision operator to balance the drive, and making these particles very sensitive to collisions. If no collisions were present, $f_1$ would diverge at resonance at a pitch angle given by a fixed $v$ and radial location. Collisions act as a diffusive operator in pitch angle
$C_{\mathrm{pas}}\{ f_1 \} \approx \nu_{\mathrm{pas}} \partial^2 f_1/\partial\lambda^2$, where
$\nu_{\mathrm{pas}} = v_{\lambda}^3 / 2 v^3 \tau_s$, allowing solutions of $f_1$ with a finite amplitude and width in $\lambda$ around the resonance. The alphas sensitive to collisions live in a boundary layer of width $\Delta\lambda$, and experience an effective collision frequency $\nu_{\mathrm{eff}} = \nu_{\mathrm{pas}}/\Delta\lambda^2$. Even if $\nu_{\mathrm{pas}}$ is small for $\alpha$ particles, due to the smallness of $\Delta\lambda$, $\nu_{\mathrm{eff}}$ requires collisions to be retained.

For trapped particles, the streaming term is removed by trajectory averaging since as the particle advances with $d\tau>0$, it traces out the same trajectory in direction $v_{\parallel}>0$ as in the opposite direction $v_{\parallel}<0$. The trapped resonance happens due to a reversal in the tangential drift for a particular pitch angle on every flux surface. The solution for the trapped is given by Catto \textit{et al.} (\citeyear{catto_merging_2023}). The perturbation of the field direction was not considered, but it averages out for the trapped. For passing particles, the ones we investigate here, the trajectory averaged parallel streaming term is finite since there is no $v_{\parallel}$ sign reversal. Instead, the resonance occurs when streaming cancels the drift term. We keep both pieces of the perturbation in \eqref{3.15}, but it is expected that field line deflection dominates over the field magnitude perturbation.

To solve \eqref{3.16} for the passing, we start by Fourier decomposing the perturbation. Assuming it must be periodic in toroidal and poloidal angles, we write
\begin{equation}
\left.
\begin{array}{l}
A_{\parallel} \\
B_{\parallel}
\end{array}
\right\}
=
\operatorname{Re} \sum_{n,m}
\left\{
\begin{array}{l}
A_{\parallel}^{nm}(\psi_h) \\
B_{\parallel}^{nm}(\psi_h)
\end{array}
\right\}
e^{i(n\zeta - m\vartheta)}
=
\operatorname{Re} \sum_{n,m}
\left\{
\begin{array}{l}
A_{\parallel}^{nm} \\
B_{\parallel}^{nm}
\end{array}
\right\}
e^{i(p^{nm}\alpha + l^{nm}\eta)} ,
\label{3.17}
\end{equation}
where $p^{nm} = (Mn - Nm)/(M - qN)$ and $l^{nm} = (q n - m)/(M - qN)$. The error fields are $B_{\parallel}^{nm}, B_{\perp}^{nm} \ll \epsilon \bar{B}_0$, with the field direction deflection estimated as $B_{\perp}^{nm} \approx 2\pi q p^{nm} A_{\parallel}^{nm}/a'$ using \eqref{2.18}. The error fields $A_{\parallel}^{nm}$ and $B_{\parallel}^{nm}$ are assumed small. Products of $A_{\parallel}^{nm} B_{\parallel}^{nm}$ will be ignored, allowing us to take $A_{\parallel}^{nm}$ and $B_{\parallel}^{nm}$ to be real and disregard their relative phase. We assume $Nm \neq Mn > 0$. We decompose $f_1$ for passing alphas by writing
\begin{equation}
f_1 =
\mathrm{Im} \sum_{n,m} f_{nm}^p(\psi_*,v,\mu,\sigma)
e^{i(n\zeta - m\vartheta)}
=
\mathrm{Im} \sum_{n,m} f_{nm}^p
e^{i( p_*^{nm} \alpha_* + l_*^{nm} \eta )},
\label{3.18}
\end{equation}
where $p_*^{nm} = (Mn - Nm)/(M - q_* N)$ and $l_*^{nm} = (q_* n - m)/(M - q_* N)$. After forming the $\alpha$ derivatives of \eqref{3.17}, we express them in $\psi_*,\eta,\alpha_*$ variables and substitute the result along with \eqref{3.18} into \eqref{3.15}. Then we multiply by $e^{-i(n'\zeta - m'\vartheta)} = e^{-i(p_*^{n'm'} \alpha_* + l_*^{n'm'}\eta)}$ and integrate $\alpha_*$ over $2\pi(M - q_*N)$ to remove the $n$ sum by requiring $M(n - n') = N(m - m')$, yielding
\[
\!\!\!\!\!\!\!\!\!\!\!\!\!\!\!
\sum_m
\left[
i\left( l_*^{nm} v_{\parallel} \bm{b}_0 \cdot \nabla \eta
+ p_*^{nm} \omega_{\alpha} \right) f_{nm}^p
- C_{\mathrm{pas}}\{ f_{nm}^p \}
\right]
e^{-i(m-m')\eta/M}
\\[-4pt]
\]
\begin{equation}
 =
- \frac{\partial f_0}{\partial\psi_p}
\sum_m p^{nm}
\left[
v_{\parallel} A_{\parallel}^{nm}
- \Omega_0^{-1}( \mu_0 B_0 + v_{\parallel}^2 ) B_{\parallel}^{nm}
\right]
e^{-i(m-m')\eta/M}.
\label{3.19}
\end{equation}

Dividing by $v_{\parallel} \bm{b}_0 \cdot \nabla \eta$ and integrating $\eta$ from $-M\pi$ to $M\pi$ leads to
\[
\sum_m
\Bigg\{
f_{nm}^p\, i \left[
l_*^{nm} \oint d\eta\, e^{-i(m-m')\eta/M}
+ p_*^{nm} \oint d\eta\, \frac{\omega_{\alpha} e^{-i(m-m')\eta/M}}{
 v_{\parallel} \bm{b}_0 \cdot \nabla \eta }
\right]
\\[-6pt]
\]
\[
- \oint d\eta\, \frac{ e^{-i(m-m')\eta/M}}{
 v_{\parallel} \bm{b}_0 \cdot \nabla \eta }\, C_{\mathrm{pas}}\{ f_{nm}^p \}
\Bigg\}
\\[-4pt]
\]
\begin{equation}
=
- \frac{\partial f_0}{\partial\psi_p}
\sum_m
\oint d\eta\, p^{nm}
[ v_{\parallel} A_{\parallel}^{nm}
- \Omega_0^{-1}( \mu_0 B_0 + v_{\parallel}^2 ) B_{\parallel}^{nm} ]
\; \frac{e^{-i(m-m')\eta/M}}{( v_{\parallel} \bm{b}_0 \cdot \nabla \eta )}.
\label{3.20}
\end{equation}

The resonance condition occurs when the streaming and drift terms cancel each other. In order for this to lead to a relevant contribution $m' = m$ is required in the first term on the left. For all other $m' \neq m$ the streaming contribution vanishes and the motion of the alpha particles is essentially non-resonant and collisionless, giving $f_{nm\neq m'}^p \to 0$. For the drift to be comparable to streaming over a relevant range of $\lambda$, we need to keep streaming small by staying in the vicinity of the rational surface with $q = m/n$. Very close to the rational surface freely passing particles with bigger $v_{\parallel}$ will be able to resonate, and farther away from the rational surface the condition will become easier to satisfy for barely passing particles with a smaller $v_{\parallel}$.

To perform these integrals, we will use $\psi,\eta,\alpha,v$ and $\lambda$ variables. The distinction between $q$ and $q_*$ is unimportant except in the resonance condition, for which a Taylor expansion will be employed in the $(q - q_*)$ term by using
\begin{equation}
(Mn - Nm) \oint d\eta\, \frac{\omega_{\alpha}}{v_{\parallel} \bm{b}_0 \cdot \nabla \eta}
= \oint d\eta\, (m - q_* n) = (m - qn) \oint d\eta + \oint d\eta\, n(q - q_*).
\label{3.21}
\end{equation}
Close to the resonant rational surface with $q = m/n$, the magnetic shear term from this Taylor expansion of $q_*$ exactly cancels the shear term from $\omega_{\alpha}$ in \eqref{3.13}. Shear is therefore unimportant for the passing particles, as they average out its effect. For the trapped alpha particles this is not the case since they experience a different shear on each leg of their bounce motion (Catto \textit{et al.} \citeyear{catto_merging_2023}). By defining
\begin{equation}
\omega_d \equiv \omega_{\alpha}
+ v_{\parallel} \bm{b}_0 \cdot \nabla \eta \frac{q_* - q}{M - qN}
\approx v_{\parallel} \bm{b}_0 \cdot \nabla \eta
\left[
q \frac{\partial}{\partial\psi_h} \Big( \frac{I v_{\parallel}}{\Omega_0} \Big)
+ \frac{\partial}{\partial\psi_h} \Big( \frac{G v_{\parallel}}{\Omega_0} \Big)
\right]
- v_{\parallel} \bm{b}_0 \cdot \nabla  \Big( \frac{K_p v_{\parallel}}{\Omega_0} \Big),
\label{3.22}
\end{equation}
the equation becomes
\[
f_{nm}^p
\left[
l^{nm} \oint d\eta
+ p^{nm} \oint d\eta\, \frac{\omega_d}{ v_{\parallel} \bm{b}_0 \cdot \nabla \eta }
\right]
+ i \oint d\eta\, \frac{C_{\mathrm{pas}}\{ f_{nm}^p \}}{ v_{\parallel} \bm{b}_0 \cdot \nabla \eta }
\\[-4pt]
\]
\begin{equation}
=
i \frac{\partial f_0}{\partial\psi_p}
p^{nm} \oint d\eta\,
\frac{
A_{\parallel}^{nm}
- \Omega_0^{-1}( \mu_0 B_0 + v_{\parallel}^2 ) B_{\parallel}^{nm}/v_{\parallel}
}{ \bm{b}_0 \cdot \nabla \eta }.
\label{3.23}
\end{equation}

The existence of this resonance produces a perturbed response which we call resonant plateau, since it drives transport with a diffusivity independent of the collisionality. For both passing and trapped, there is another piece of the perturbed distribution function that can ensure $f_1$ vanishes at the trapped-passing boundary. This causes what is known as $\sqrt{\nu}$ transport, since the diffusivity associated with it has such a collisionality dependence, but it produces a smaller contribution to energy losses as discussed in Appendix~\ref{appendixB}.

    \subsection{Solution to the $f_1$ equation}\label{subsection33}

It is convenient to consider a large aspect ratio QS stellarator with an inverse aspect ratio $\epsilon = r/R \ll 1$. Its background magnetic field amplitude can be described as
\begin{equation}
B_0 = \bar{B}_0 \left[ 1 - \epsilon \cos(\eta) \right].
\label{3.24}
\end{equation}

We assume $I \approx \bar{B}_0 R$ and a driving toroidal field much bigger than the poloidal field $qI \gg G$ and $MI \gg NG$, and also much larger than the radial field $K_p \ll 2\pi q/a'$. We employ
$d\psi_h = (M - qN)\bar{B}_0 a' \, dr / 2\pi q$ and $\bm{b}_0 \cdot \nabla  \eta \approx (M - qN)/qR$ to approximate the drift as
\begin{equation}
\omega_d \approx q v_{\parallel} \bm{b}_0 \cdot \nabla \eta \,
\frac{\partial}{\partial\psi_h} \Big( \frac{I v_{\parallel}}{\Omega_0} \Big)
\approx \frac{2\pi q v_{\parallel} \bar{B}_0}{\bar{\Omega}_0 a'} \,
\frac{\partial ( v_{\parallel}/B_0 )}{\partial r}
\approx \frac{ \pi q v^2 (2 - \lambda) \cos\eta}{ \bar{\Omega}_0 a' R }.
\label{3.25}
\end{equation}
Using \eqref{3.25} we verify that for the drift $\omega_d \sim q v^2/\bar{\Omega}_0 a' R$ to be comparable to the parallel streaming $k_{\parallel} v_{\parallel}$, with $k_{\parallel} = (qn - m)/qR$, we need to be somewhat close to the resonant rational surface $q = m/n$ since
\begin{equation}
\frac{|q - m/n|}{q} \sim \frac{qv}{n\bar{\Omega}_0 a'} \sim \frac{q\rho}{nr} \ll 1.
\label{3.26}
\end{equation}

To implement the effects of the trajectory average and evaluate the $\eta$ integrals in the equation for $f_{nm}^p$ it is convenient to define
$\lambda = k^2 / [ (1-\epsilon)k^2 + 2\epsilon ]$ and write
\begin{equation}
\xi = |v_{\parallel}|/v = \sqrt{1 - \lambda(1 - \epsilon\cos\eta)}
= (\sqrt{2\epsilon\lambda}/k)\sqrt{ 1 - k^2 \sin^2(\eta/2) },
\label{3.27}
\end{equation}
where $k$ is a useful variable for the pitch angle that goes from 0 for freely passing to 1 for barely passing particles. Unlike $\lambda$, $k$ is not a true adiabatic invariant because of its $\epsilon$ dependence. However, for our purposes $k$ can be treated as a constant of the motion provided the width of the collisional boundary layer is larger than the width of any island structure created by the perturbation, as discussed in section \ref{section4}. The following integrals are required
\begin{equation}
\int_{-\pi}^{\pi} d\eta / \xi =
4 \sqrt{ (1-\epsilon)k^2 + 2\epsilon } \, K(k) / \sqrt{2\epsilon},
\label{3.28}
\end{equation}
\begin{equation}
\int_{-\pi}^{\pi} d\eta \, \xi =
4 \sqrt{2\epsilon}\, E(k)/\sqrt{ (1-\epsilon)k^2 + 2\epsilon },
\label{3.29}
\end{equation}
\begin{equation}
\int_{-\pi}^{\pi} d\eta \, \cos\eta / \xi
=
4 \sqrt{ (1-\epsilon)k^2 + 2\epsilon } \,
[ 2E(k) - (2 - k^2)K(k) ] / \sqrt{2\epsilon}k^2,
\label{3.30}
\end{equation}
where $K(k)$ and $E(k)$ are complete elliptic integrals of the first and second kind, respectively. These expressions allow us to calculate the drift term
\begin{equation}
\int_{-\pi}^{\pi} d\eta \, \frac{\omega_d}{\xi}
= -\frac{4\pi q v^2 F}{\bar{\Omega}_0 a' R \sqrt{2\epsilon}}
\equiv -2\pi \omega_{d0} F,
\label{3.31}
\end{equation}
where $\omega_{d0} = 2 q v^2/\bar{\Omega}_0 a' R\sqrt{2\epsilon}$ and
\begin{equation}
F \equiv \frac{-[(1-2\epsilon) k^2 + 4\epsilon]}{ k^2 \sqrt{(1-\epsilon)k^2 + 2\epsilon} }\, G > 0,
\label{3.32}
\end{equation}
with $G = [2E(k) - (2 - k^2)K(k)]$. Similarly, using
$\left.\partial\lambda/\partial k\right|_{\epsilon} = 4\epsilon k / [ (1-\epsilon)k^2 + 2\epsilon ]^2$,
the collision operator term is
\[
\int_{-\pi}^{\pi} d\eta \frac{\; C_{\mathrm{pas}}\{ f_{nm}^p \}}{\xi}
=
\frac{2 v_{\lambda}^3}{ \tau_s v^3 }
\frac{\partial}{\partial\lambda} \left[ (\lambda \int_{-\pi}^{\pi} d\eta \, \xi \, )
\frac{\partial f_{nm}^p}{\partial\lambda } \right]
\]
\[
=
\frac{ 2 v_{\lambda}^3 [ (1-\epsilon)k^2 + 2\epsilon ]^2 }
{ \tau_s v^3 (2\epsilon)^{3/2} k }
\; \frac{\partial}{\partial k} \left[
k \sqrt{(1-\epsilon)k^2 + 2\epsilon}\, E(k) \,
\frac{\partial f_{nm}^p}{\partial k}
\right]
\]
\begin{equation}
\approx
\frac{2 v_{\lambda}^3 [ (1-\epsilon)k^2 + 2\epsilon ]^{5/2} E(k)}{\tau_s v^3 (2\epsilon)^{3/2}}
\; \frac{\partial^2 f_{nm}^p}{\partial k^2}
\equiv \frac{ \sqrt{2}\, \nu_{\mathrm{pas}} C_{\nu}(k) }{\epsilon^{3/2}}
\; \frac{\partial^2 f_{nm}^p}{\partial k^2},
\label{3.33}
\end{equation}
where we defined
$C_{\nu}(k) = [ (1-\epsilon)k^2 + 2\epsilon ]^{5/2} E(k)$.

We substitute the preceding into the kinetic equation to obtain
\begin{equation}
f_{nm}^p \Big[ \sigma(qn-m) \frac{v}{qR} - p^{nm} \omega_{d0} F \Big]
+ i \frac{ \nu_{\mathrm{pas}} C_{\nu}(k) }{ \pi (2\epsilon^3)^{1/2} }
\frac{ \partial^2 f_{nm}^p}{\partial k^2 }
=
i p^{nm} v \frac{\partial f_0}{\partial\psi_p}
[ \sigma A_{\parallel}^{nm} + a' a_{\parallel}(k) B_{\parallel}^{nm} ],
\label{3.34}
\end{equation}
%
with the prefactor of the $B_{\parallel}^{nm}$ term defined as
\begin{equation}
a_{\parallel}(k)
=
-\frac{2\mu_0 B_0 \sqrt{(1-\epsilon)k^2 + 2\epsilon}}{\sqrt{2\epsilon}\, \pi \Omega_0 v a'}
\;
\left[
K(k)
+ \frac{2\epsilon v^2 E(k)/\mu_0 B_0}
{ (1-\epsilon)k^2 + 2\epsilon }
\right].
\label{3.35}
\end{equation}

\begin{figure}
  \centering
  \includegraphics[width=0.9\textwidth]{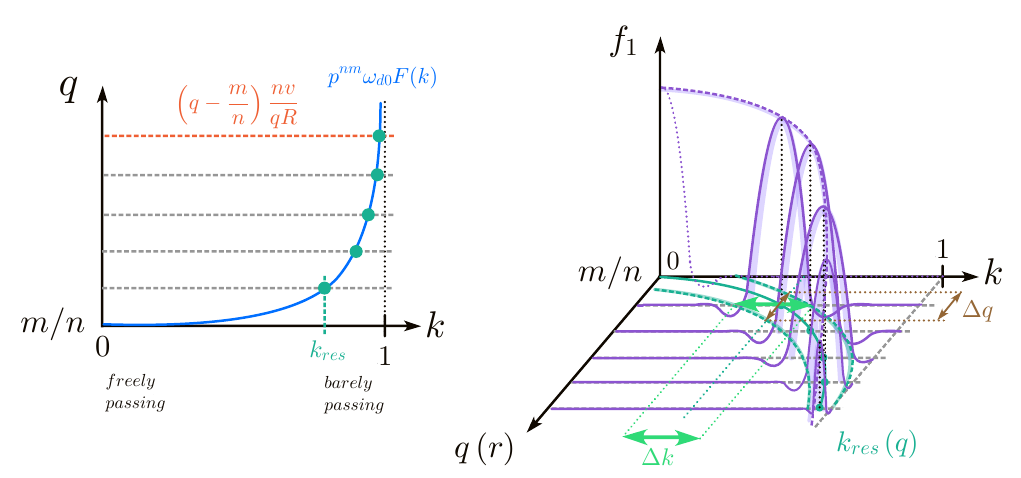}
  \caption{Resonance condition (left) and schematic of the collisional boundary layer perturbed solution around the resonance (right) as we move away from the rational surface.}
  \label{figure1}
\end{figure}

Given that the drift term doesn’t change sign, the resonance condition between streaming and drift will happen for $\sigma = 1$ when $q > m/n$, and for $\sigma = -1$ when $q < m/n$. We assume $p^{nm} > 0$. Then, the resonance occurs at $k_{\mathrm{res}}$ when $q \approx m/n$ and with $p^{nm} \approx n > 0$ so that
\begin{equation}
|qn - m| (v/qR) = p^{nm} \omega_{d0} F(k_{\mathrm{res}}).\\[-4pt]
\\[8pt]
\label{3.36}
\end{equation}
Expanding gives $F = \sqrt{2\epsilon} \, \pi k^2/8 + O(k^4)$ for $k \to 0$, while
$F \approx - \ln \sqrt{1-k}$ diverges as $k \to 1$. Therefore, $F(k)$ samples all real positive values, and a $k_{\mathrm{res}}$ can in principle be found for any value of $q$, as depicted in Figure \ref{figure1} (left). The dashed horizontal lines represent streaming and intersect the drift term blue curve to give the resonant $k_{\mathrm{res}}$.

The local approximation to the streaming plus drift term near the resonance can be calculated by a Taylor expansion about $k_{\mathrm{res}}$ as
\begin{equation}
F(k)\approx C_0 + C_1 (k - k_{\mathrm{res}}) ,
\label{3.37}
\end{equation}
where $C_0$ ensures the cancellation between drift and streaming for every value of $q(\psi_h)$ to lowest order. It defines the resonant pitch angle $k_{\mathrm{res}}$, which has to satisfy
\begin{equation}
C_0 = F(k_{\mathrm{res}}) =
\left. \frac{ -[(1-2\epsilon) k^2 + 4\epsilon] }{ k^2 \sqrt{ (1-\epsilon)k^2 + 2\epsilon } } \, G \right|_{k = k_{\mathrm{res}}}
= \frac{|qn - m|}{\omega_{d0} p^{nm}} \frac{v}{qR} .
\label{3.38}
\end{equation}
The coefficient $ C_1=\partial F/ \partial k |_{\epsilon, k = k_{\mathrm{res}}} $ is evaluated to be
\begin{equation}
\! C_1 = \!
\left.
\frac{ (1\!-\!2\epsilon)k^2 + 4\epsilon }{ \sqrt{ (1\!-\!\epsilon)k^2 + 2\epsilon } } \frac{dK(k)}{dk}
\!+\!
\frac{ (1\!-\!\epsilon)(1\!-\!2\epsilon)k^4 + 12\epsilon(1\!-\!\epsilon)k^2 + 16\epsilon^2 }
     { k^3 [ (1\!-\!\epsilon)k^2 + 2\epsilon ]^{3/2} }
\, G
\right|_{k = k_{\mathrm{res}}} ,
\label{3.39}
\end{equation}
where $dE(k)/dk = [E(k)\!-\!K(k)]/k$ and $dK(k)/dk = [E(k)\!-\!(1\!-\!k^2)K(k)]/[k(1\!-\!k^2)]$ are used to obtain $dG/dk = -k^2 dK(k)/dk$. Given the slow variation of $C_\nu(k)$ and $a_{\parallel}(k)$ within the narrow collisional boundary layer width, estimated in section \ref{section4}, we can assume $C_\nu(k)\approx C_\nu(k_{\mathrm{res}})$ and $a_{\parallel}(k)\approx a_{\parallel}(k_{\mathrm{res}})$. In addition, for $k_{\mathrm{res}}^2 \sim 2\epsilon$ we notice
$a' a_{\parallel}(k) B_{\parallel}^{nm}/A_{\parallel}^{nm} \sim q n v B_{\parallel}^{nm}/\bar{\Omega}_0 r B_{\perp}^{nm}$,
suggesting the $A_{\parallel}^{nm}$ term will dominate when $qnv/\bar{\Omega}_0 r \ll 1$.

If we define $A_{nm} = A_{\parallel}^{nm} + \sigma a' a_{\parallel}(k_{\mathrm{res}}) B_{\parallel}^{nm}$, we can write the kinetic equation as
\begin{equation}
p^{nm} \omega_{d0} C_1 (k_{\mathrm{res}} - k) f_{nm}^p
+ i \frac{ \nu_{\mathrm{pas}} C_\nu }{ \pi (2\epsilon^3)^{1/2} }
\frac{\partial^2 f_{nm}^p}{\partial k^2}
=
i \sigma p^{nm} v \frac{\partial f_0}{\partial \psi_p} A_{nm} .
\label{3.40}
\end{equation}
By defining $\Xi (k_{\mathrm{res}} - k) = Z$, with
$
\Xi = \left[{ \pi (2\epsilon)^{3/2} p^{nm} \omega_{d0} C_1 / C_\nu \nu_{\mathrm{pas}} } \right]^{1/3},
$
the previous equation becomes an inhomogeneous Airy equation
\begin{equation}
\frac{\partial^2 f_{nm}^p}{\partial Z^2}
- i Z f_{nm}^p
=
\sigma v \frac{\Xi}{\omega_{d0} C_1}
\frac{\partial f_0}{\partial \psi_p}
A_{nm} .
\label{3.41}
\end{equation}
The most relevant piece of the solution to this equation is the inhomogeneous part, which corresponds to a Su–Oberman solution (Su and Oberman \citeyear{su_collisional_1968})
\begin{equation}
\Upsilon_{\mathrm{SO}}(Z)
=
\int_0^\infty d\tau\, e^{-i Z \tau - \tau^3/3} ,
\label{3.42}
\\[-4pt]
\end{equation}
giving
\begin{equation}
f_{nm}^p
=
- \frac{\sigma v \Xi}{\omega_{d0} C_1}
\frac{\partial f_0}{\partial \psi_p}
A_{nm}
\Upsilon_{\mathrm{SO}}(Z) .
\label{3.43}
\end{equation}
A depiction of this solution is shown on the right of Figure \ref{figure1}. Closer to the rational surface, freely passing particles with $k_{\mathrm{res}}\to 0$ are more easily in resonance. However, when we are very close to $k_{\mathrm{res}} \approx 0$ the approximation in \eqref{3.33} cannot be made and the perturbed distribution function becomes less localized in $k$. As we move away from $q=m/n$, $k_{\mathrm{res}}$ increases, and when we move far enough away it is the barely passing with $k_{\mathrm{res}} \to 1$ that resonate. As $k_{\mathrm{res}} \to 1$, the homogeneous piece of the solution can come into play to ensure that at the trapped–passing boundary $f_{nm}^p(k=1)=0$, which makes $f_1$ smaller. As shown in Appendix~\ref{appendixB}, this piece has the form of an Airy function and produces $\sqrt{\nu}$ transport, but of smaller magnitude than the resonant plateau transport. When $k_{\mathrm{res}}$ is extremely close to 1, it can become relevant and comparable to the inhomogeneous term as both get pushed onto the trapped–passing boundary, but their solutions and resulting transport will be very small.
    
    \subsection{Passing alpha particles energy flux}\label{subsection34}

Now that we have obtained the solution for $f_1$, we can calculate the heat flux that this distribution drives in the transit averaged quasilinear equation for $f_0$
\begin{equation}
\frac{S\delta(v-v_0)}{4\pi v^2}
+\frac{1}{\tau_s v^2}\frac{\partial}{\partial v}\big[(v^3+v_c^3)f_0\big]
+\frac{2 v_\lambda^3}{\tau_s v^5 \oint_\alpha d\tau}
\frac{\partial}{\partial\lambda}\!\left[\lambda \Big(\!\oint_\alpha d\tau\, v_{\parallel}^2\Big)
\frac{\partial f_0}{\partial\lambda}\right]
=-\mathcal{L}\{f_0\} ,
\label{3.44}
\end{equation}
where as seen from \eqref{3.10}, with $f_1$, $A_{\parallel}$ and $B_{\parallel}$ inserted, the losses enter through
\begin{equation}
\mathcal{L}\{f_0\}
=-\frac{1}{2\oint d\tau}\frac{\partial}{\partial\psi_p}
\left[\sum_{n,m} p^{nm}\!\oint d\tau\, v_{\parallel} A_{nm}\,\mathrm{Re}(f_{nm}^p)\right] ,
\label{3.45}
\end{equation}
with $\langle \sin^2(p^{nm}\alpha + l^{nm}\eta)\rangle = 1/2$ and only the terms with the same poloidal and toroidal numbers contributing to lowest order. Rewriting \eqref{3.45} using $\oint d\tau\, v_{\parallel}
= \sigma\int_{-\pi}^{\pi}{d\eta}/{\bm{b}_0\!\cdot\nabla \eta}\approx {2\pi\sigma q R}/{(M-qN)}$ and ${\partial}/{\partial\psi_p}
\approx
({2\pi q}/{\bar{B}_0 a'}){\partial}/{\partial r}$
leads to
\begin{equation}
\mathcal{L}\{f_0\}
= -\frac{2\pi^2 q\sigma}{\bar{B}_0 a' \oint d\tau}
\frac{\partial}{\partial r}
\left[
\frac{qR}{M-qN}
\sum_{n,m} p^{nm} A_{nm} \mathrm{Re}(f_{nm}^p)
\right].
\label{3.46}
\end{equation}

To calculate the divergence of the energy flux, we use
$d^3v = \pi v^2 dv\, d\lambda\, B_0/(\xi \bar{B}_0)$ and the flux surface average, as
\begin{equation}
\begin{aligned}
&\mathcal{Q} \equiv \frac{1}{2}M_\alpha
\Big\langle \int d^3v\, v^2\, \mathcal{L}\{f_0\}\Big\rangle
= \frac{M_\alpha (M-qN)}{4qR}
\int_0^\infty dv\, v^5
\int_0^{1-\epsilon} d\lambda\, \mathcal{L}\{f_0\} {\oint}
 d\tau 
\\&
= -\frac{\sigma\pi^2 M_\alpha (M-qN)}{2 qR\bar{B}_0 a'}
\frac{\partial}{\partial r}
\left[
\frac{qR}{M-qN}
\sum_{n,m} p^{nm}
\int_0^\infty dv\, v^5
\int_0^{1-\epsilon} d\lambda\,
A_{nm}\,\mathrm{Re}(f_{nm}^p)
\right]
\\&
= \frac{\pi^3 M_\alpha (M-qN)}{\bar{B}_0^2 R a'}
\frac{\partial}{\partial r}
\left[
\sum_{n,m}
\frac{q^2 R\, p^{nm} A_{nm}^2}{a'(M-qN)}
\int_0^\infty dv\, \frac{v^6 \Xi}{\omega_{d0} C_1}
\frac{\partial f_0}{\partial r}
\int_0^{1-\epsilon} d\lambda
\int_0^\infty d\tau\, e^{-{\tau^3}/{3}}\cos(Z\tau)
\right].
\label{3.47}
\end{aligned}
\end{equation}
We perform the pitch angle integral first, taking advantage of the delta function behavior of the Su–Oberman solution since $\Xi\gg 1$. The function
$\partial\lambda/\partial k = 4\epsilon k/[(1-\epsilon)k^2+2\epsilon]^2$
varies slowly in comparison and can be assumed constant up to $O(\Xi^{-2})$ corrections, as described in Appendix~\ref{appendixC}
\begin{equation}
\int_0^{1-\epsilon} d\lambda
\int_0^\infty d\tau\, e^{-\tau^3/3}\cos(Z\tau)
\approx
\frac{\pi}{\Xi}
\left.\frac{\partial\lambda}{\partial k}\right|_{\epsilon,k=k_{\mathrm{res}}}.
\label{3.48}
\end{equation}
Assuming a background slowing down tail distribution, which satisfies
$\partial f_0/\partial r \approx [4\pi (v^3+v_c^3)]^{-1} \partial(S\tau_s)/\partial r$,
and defining
${\omega}_0 = 2qv_0^2/\bar{\Omega}_0 a' R \sqrt{2\epsilon}$,
we obtain
\begin{equation}
\mathcal{Q} = \frac{\pi^3 M_\alpha (M-qN)}{8a'}
\frac{\partial}{\partial r}
\left[
\frac{a' v_0^4 q^2}{{\omega}_0 (M-qN)}
\left(
\sum_{n,m}
\frac{ p^{nm} A_{nm}^2 }
     { C_1 \bar{B}_0^2 (a')^2 }
\left.
\frac{\partial\lambda}{\partial k}\right|_{\epsilon, k=k_{\mathrm{res}}}
\right)
\frac{\partial (S\tau_s)}{\partial r}
\right].
\label{3.49}
\end{equation}
by using $\int_0^{v_0} dv\, {v^4}/{(v^3 + v_c^3)} \approx {v_0^2}/{2}.$

To obtain the energy flux $\Gamma$, we use the QS field approximation for the flux surface average of the divergence,
$\mathcal{Q} \approx - [(M-qN)/a']\, \partial [a'\Gamma/(M-qN)]/\partial r$, to find
\begin{equation}
\Gamma
=
-\sum_{n,m}
\frac{\pi^3 M_\alpha}{8}
\frac{ v_0^4 q^2 }{{\omega}_0}
\left(
\sum_{n,m}
\frac{ p^{nm} A_{nm}^2 }
     { C_1 \bar{B}_0^2 (a')^2 }
\left.\frac{\partial\lambda}{\partial k}\right|_{\epsilon,k=k_{\mathrm{res}}}
\right)
\frac{\partial(S\tau_s)}{\partial r}.
\label{3.50}
\end{equation}
Introducing the alpha particle density for the slowing down tail,
$n_\alpha \approx S\tau_s \ln(v_0/v_c)$,
the energy diffusivity is defined by
$\Gamma = -(M_\alpha v_0^2/2) D\, \partial n_\alpha/\partial r$, which gives
\begin{equation}
D
=
\frac{\pi^3 v_0^2 q^2}{4 {\omega}_0 \ln(v_0/v_c)}
\sum_{n,m}
\left(
\frac{ p^{nm} }{ C_1 }
\left.\frac{\partial\lambda}{\partial k}\right|_{\epsilon,k=k_{\mathrm{res}}}
\right)
\left(
\frac{A_{nm}}{\bar{B}_0 a'}
\right)^2 .
\label{3.51}
\end{equation}
Recalling the relation between $B_{\perp}^{nm}$ and $A_{\parallel}^{nm}$ after \eqref{3.17}, and defining
$a_{\perp} = 1/2\pi q p^{nm}$, we arrive at our final expression

\begin{equation}
D
=
\frac{\pi v_0^2}{16{\omega}_0 \ln(v_0/v_c)}
\sum_{n,m}
\frac{1}{p^{nm} C_1}
\left.\frac{\partial\lambda}{\partial k}\right|_{\epsilon,k=k_{\mathrm{res}}}
\left[
\left(\frac{B_{\perp}^{nm}}{\bar{B}_0}\right)^2
+
\frac{ a_\parallel^2(k_{\mathrm{res}}) }{ a_{\perp}^2 }
\left(\frac{B_{\parallel}^{nm}}{\bar{B}_0}\right)^2
\right].
\label{3.52}
\end{equation}
On a flux surface of safety factor $q$, the major contributions to be retained in the sum are those with $n$ and $m$ such that $m/n \approx q$. 

We can now estimate alpha energy losses with \eqref{3.52}. Letting the minor radius $r_0$ label the flux surface just before the separatrix, where the aspect ratio expansion should remain valid, the relevant parameter to evaluate and keep small is $D\tau_s / r_0^2$. If a large fraction of alpha energy is lost by a radial length $r_0$ before alphas can slow down in a time $\tau_s$, then wall damage and/or inadequate alpha heating will occur. In sections \ref{section4} and \ref{section5}, we discuss and calculate this transport and the validity constraints of the model.

\section{Phenomenological estimates and validity of the model}\label{section4}

To gain understanding of the physical picture and test the limits of the model validity from simple estimates, we write down \eqref{3.40} using the pitch angle variable $\lambda$ as
\begin{equation}
i p^{nm} \, \frac{\omega_{d0} C_1}{\partial\lambda/\partial k}\, (\lambda - \lambda_{\mathrm{res}}) f_{nm}^p
+ \frac{1}{\pi} \frac{v_\lambda^3}{v^3 \tau_s}
\frac{\partial}{\partial\lambda}
\left[
\frac{4\lambda \sqrt{2\epsilon} \, E(k)}{\sqrt{(1-\epsilon)k^2 + 2\epsilon}}
\frac{\partial f_{nm}^p}{\partial\lambda}
\right]
=
\sigma p^{nm} v \, \frac{\partial (f_0 + f_{nm}^p)}{\partial\psi_p} A_{nm} .
\label{4.1}
\end{equation}
Note that the $\partial f_{nm}^p/\partial\psi_h$ term has been kept for generality, and permits us to estimate when linearizing the $f_1$ equation is allowed. For the purposes of these estimates, we will denote $f_{nm}^p$ as $f_1$.

The streaming plus drift term can be estimated as
\begin{equation}
i p^{nm} \, \frac{\omega_{d0} C_1}{\partial\lambda/\partial k}\, (\lambda - \lambda_{\mathrm{res}}) f_{nm}^p
\sim \tilde{\omega}_d \, \Delta\lambda \, f_1 ,
\label{4.2}
\end{equation}
with $\tilde{\omega}_d = p^{nm} \omega_0 \left[ C_1 / (\partial\lambda/\partial k)|_\epsilon \right]_{k = k_{\mathrm{res}}}$ and $\Delta\lambda \approx \lambda - \lambda_{\mathrm{res}}$ is associated with the pitch angle width of the collisional boundary layer. The collision operator term can be approximated as
\begin{equation}
\frac{1}{\pi} \frac{v_\lambda^3}{v^3 \tau_s}
\frac{\partial}{\partial\lambda}
\left[
\frac{4\lambda \sqrt{2\epsilon} \, E(k)}{\sqrt{(1-\epsilon)k^2 + 2\epsilon}}
\frac{\partial f_{nm}^p}{\partial\lambda}
\right]
\sim
\frac{\nu_{\mathrm{pas}}}{\pi} \,
\frac{8\sqrt{2\epsilon} \, k^2 E(k)}{[(1-\epsilon)k^2 + 2\epsilon]^{3/2}}
\, \frac{\partial^2 f_{nm}^p}{\partial\lambda^2}
\sim
\frac{\tilde{\nu}_{\mathrm{pas}}}{\Delta\lambda^2} \, f_1 ,
\label{4.3}
\end{equation}
with
$\tilde{\nu}_{\mathrm{pas}} = 8\sqrt{2\epsilon}\, k_{\mathrm{res}}^2 E(k_{\mathrm{res}})\, \nu_{\mathrm{pas}} /
\big\{\pi[(1-\epsilon)k_{\mathrm{res}}^2 + 2\epsilon]^{3/2}\big\}$.
The drive term is estimated as
\begin{equation}
\sigma p^{nm} v \, \frac{\partial (f_0 + f_{nm}^p)}{\partial\psi_h} A_{nm}
\sim
V \left( \frac{\partial f_0}{\partial r} + \frac{\partial f_1}{\partial r} \right) ,
\label{4.4}
\end{equation}
where $V = 2\pi q p^{nm} \sigma v_0 A_{nm}/\bar{B}_0 a' = v_0 \sigma B_{nm}/\bar{B}_0$ and
$B_{nm} = B_{\perp}^{nm} + [\sigma a_{\parallel}(k_{\mathrm{res}})/a_{\perp}] B_{\parallel}^{nm}$.
We can then schematically approximate the \eqref{4.1} equation by
\begin{equation}
\tilde{\omega}_d \, \Delta\lambda \, f_1
+ \frac{\tilde{\nu}_{\mathrm{pas}}}{\Delta\lambda^2} \, f_1
\sim
V \left( \frac{\partial f_0}{\partial r} + \frac{\partial f_1}{\partial r} \right) .
\label{4.5}
\end{equation}

The collision operator term becomes comparable to the streaming plus drift term around the resonance. This produces a collisional boundary layer for each flux surface near the rational surface $q = m/n$ around its resonant pitch angle $\lambda_{\mathrm{res}}$ of width $(\Delta\lambda)_\nu$, estimated as
\begin{equation}
(\Delta\lambda)_\nu \, \tilde{\omega}_d \sim \frac{\tilde{\nu}_{\mathrm{pas}}}{(\Delta\lambda)_\nu^2}
\quad\Rightarrow\quad
(\Delta\lambda)_\nu \sim \Big( \frac{\tilde{\nu}_{\mathrm{pas}}}{\tilde{\omega}_d} \Big)^{1/3} .
\label{4.6}
\end{equation}
Since $(\Delta\lambda)_\nu \ll 1$, the effective collisionality $\nu_{\mathrm{eff}}$ in this layer is enhanced, and becomes
\begin{equation}
\nu_{\mathrm{eff}} \sim \frac{\tilde{\nu}_{\mathrm{pas}}}{(\Delta\lambda)_\nu^2}
\sim \tilde{\nu}_{\mathrm{pas}}^{1/3} \tilde{\omega}_d^{2/3} .
\label{4.7}
\end{equation}
Assuming the fraction of particles undergoing this process is $\mathcal{F} \sim (\Delta\lambda)_\nu$, the diffusivity can be estimated as
\begin{equation}
D \sim \mathcal{F} \frac{V^2}{\nu_{\mathrm{eff}}}
\sim \frac{(\Delta\lambda)_\nu^3 V^2}{\tilde{\nu}_{\mathrm{pas}}}
\sim \frac{V^2}{\tilde{\omega}_d} .
\label{4.8}
\end{equation}
which is consistent with the result in \eqref{3.52} and confirms that the resulting diffusivity has a resonant plateau behavior, exhibiting no explicit dependence on the collision frequency.

We now check the range of validity of the model. To linearize the $f_1$ equation the nonlinear term
$V \partial f_1/\partial r \sim V f_1/\Delta r$ has to remain small, with $\Delta r$ the radial scale of variation of $f_1$. We use the resonance condition \eqref{3.36} to estimate the radial width of $f_1$. Since $f_1$ is centered around $\lambda_{\mathrm{res}}(\epsilon,k_{\mathrm{res}})$, when we shift by $\Delta r$ this causes a $\Delta\lambda_{\mathrm{res}}$ given implicitly by
\begin{equation}
\Delta r \, \frac{\partial q}{\partial r} \,
\frac{\partial}{\partial q}
\left( |qn-m| \, \frac{M-qN}{q^2} \right)
=
\frac{v_0 (Mn-Nm)}{\bar{\Omega}_0 \pi R \sqrt{2}}
\, \Delta \Big(\frac{F}{\epsilon^{3/2}} \Big) ,
\label{4.9}
\end{equation}
\[\\[-20pt]\]
and
\begin{equation}
\Delta F
=
\left. \frac{\partial F}{\partial k}\right|_{\epsilon} \Delta k_{\mathrm{res}}
+ \left. \frac{\partial F}{\partial\epsilon}\right|_{k} \Delta\epsilon
=
\frac{\left.
{\partial F}/{\partial k}\right|_{\epsilon}}{ \left.{\partial\lambda}/{\partial k}\right|_{\epsilon}}
\, \Delta\lambda_{\mathrm{res}}
+
\left(
\left.\frac{\partial F}{\partial k}\right|_{\epsilon}
\left.\frac{\partial k}{\partial\epsilon}\right|_{\lambda}
+
\left.\frac{\partial F}{\partial\epsilon}\right|_{k}
\right) \Delta\epsilon ,
\label{4.10}
\end{equation}
where we used
$\Delta k_{\mathrm{res}}
=
\left.{\partial k}/{\partial\epsilon}\right|_{\lambda} \Delta\epsilon
+
\left. {\partial k}/{\partial\lambda}\right|_{\epsilon} \Delta\lambda_{\mathrm{res}}$.
The exact function $d\lambda/dr$ that relates $\Delta\lambda_{\mathrm{res}}$ and $\Delta r$ is given in Appendix~\ref{appendixD}. The collisional boundary layer radial width $(\Delta r)_\nu$ can be estimated as
\begin{equation}
(\Delta r)_\nu = (\Delta\lambda)_\nu / (d\lambda/dr) .
\label{4.11}
\end{equation}

The nonlinear term attempts to create an island with radial width $(\Delta r)_{\mathrm{is}}$ and an associated shift
$(\Delta\lambda)_{\mathrm{is}} = (d\lambda/dr)(\Delta r)_{\mathrm{is}}$. The linearized treatment remains valid until the island width becomes larger than the collisional boundary layer width. We can estimate the radial island width by balancing the radially perturbed streaming plus drift term with the nonlinear term
\begin{equation}
\frac{n v }{qR}  \, \frac{\partial q}{\partial r} (\Delta r)_{\mathrm{is}}
\sim V/(\Delta r)_{\mathrm{is}} ,
\label{4.12}
\end{equation}
%
to find
\begin{equation}
(\Delta r)_{\mathrm{is}} \sim \bigg[\frac{qR V}{n v \, \partial q/\partial r} \bigg]^{1/2} .
\label{4.13}
\end{equation}

As we require $(\Delta\lambda)_\nu > (\Delta\lambda)_{\mathrm{is}} \sim V/\tilde{\omega}_d (\Delta r)_{\mathrm{is}}$ and
$(\Delta r)_\nu > (\Delta r)_{\mathrm{is}}$, the linearization condition is
\begin{equation}
V < \tilde{\omega}_d (\Delta\lambda)_\nu (\Delta r)_\nu .
\label{4.14}
\end{equation}
This condition is equivalent to requiring the nonlinear term to be smaller than the drift plus streaming term in the collisional layer $V f_1/(\Delta r)_\nu < \tilde{\omega}_d (\Delta\lambda)_\nu f_1$. Using $B_{\parallel}^{nm} \sim B_{\perp}^{nm} \sim \delta B$, then
$V = (1 + a_{\parallel}/a_{\perp}) v \delta B/\bar{B}_0$, and the preceding becomes a constraint on the perturbation amplitude
%
\begin{equation}
\delta B < \frac{\tilde{\omega}_d (\Delta\lambda)_\nu (\Delta r)_\nu}{v(1+a_{\parallel}/a_{\perp})} \bar{B}_0
\equiv \delta B_{\max}^{\mathrm{gen}} .
\label{4.15}
\end{equation}
This linearization condition must be satisfied to evaluate the flux quasilinearly. It imposes a constraint on the model validity, as evaluated in section \ref{section5}. It also allows us to employ the unperturbed quasisymmetric particle orbits. Interestingly, this treatment only requires flux surfaces to be well defined within $(\Delta r)_\nu$.

When $k_{\mathrm{res}}\to 0$, the $f_1$ equation can change and make $f_1$ less localized in $\lambda$, resulting in a larger $\Delta r$. This indicates that our estimate of $\delta B_{\max}^{\mathrm{gen}}$ in \eqref{4.15} might break down be excessively small for $k_{\mathrm{res}}\to 0$. For this reason, we also consider an alternative constraint
$\delta B_{\max}^{k\to 1} = \delta B_{\max}^{\mathrm{gen}}(k_{\mathrm{res}}\to 1)$. As discussed in Appendix \ref{appendixD}, we expect the effect of the nonlinearity to flatten the transport curves in section \ref{section5}, and it is reasonable to expect this effect to take place somewhere between the values given by the $\delta B_{\max}^{\mathrm{gen}}$ and $\delta B_{\max}^{k\to 1}$ constraints.

Our treatment is consistent provided the resonant plateau losses are small in a slowing down time, requiring $D \sim V^2/\tilde{\omega}_d < r_0^2/\tau_s$, which gives
\begin{equation}
\delta B < \frac{1}{1+a_{\parallel}/a_{\perp}}
\frac{r_0}{v} \Big(\frac{\tilde{\omega}_d}{\tau_s}\Big)^{1/2} \bar{B}_0
\equiv \delta B_{\max}^{100} .
\label{4.16}
\end{equation}
It is shown in section \ref{section5} that this condition is less restrictive than the linearization constraint, since generally
$\delta B_{\max}^{\mathrm{gen}} / \delta B_{\max}^{100}
= (\tilde{\omega}_d \tau_s)^{1/2} (\Delta\lambda)_\nu (\Delta r)_\nu/r_0 < 1$.
Therefore, the model can self-consistently predict significant energy losses, as long as they are not close to $100\%$. When the linearization constraint is not satisfied, the radial energy loss is expected to be reduced by a smoothing of the transport curves (Catto \citeyear{catto_collisional_2025}).

\section{Estimates of alpha energy transport}\label{section5}
    \subsection{Energy diffusivity of a baseline case stellarator}\label{subsection51}

\begin{figure}
    \centering
    \includegraphics[width=0.85\linewidth]{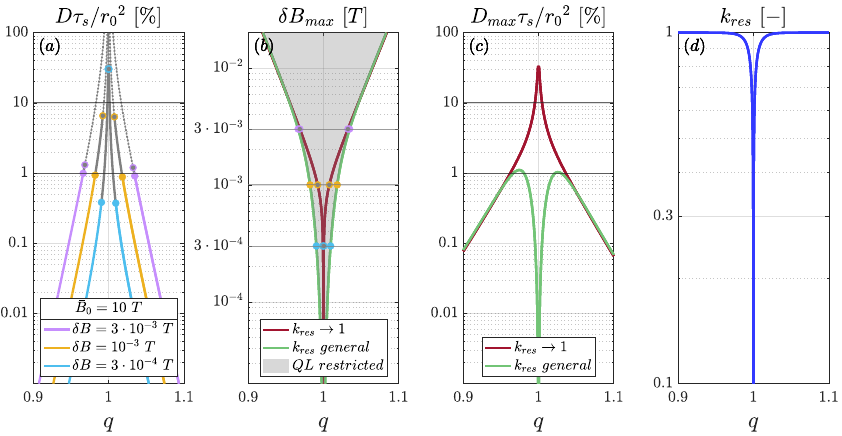}
    \caption{Transport results for the mode $m=n=3$ versus safety factor $q$. (a) Transport levels for different perturbation amplitudes. (b) Maximum perturbation allowed by the linearization condition. (c) Transport for the maximum allowable perturbations. (d) Resonant pitch angle.}
    \label{figure2}
\end{figure}

To make the figures in this section we assume the following reasonable parameters for a future stellarator: an average magnetic field $\bar{B}_0 = 10~\mathrm{T}$, major radius $R = 13~\mathrm{m}$, minor radius $r_0 = 1.3~\mathrm{m}$, global shear $s = -q^{-1} r\,\partial q/\partial r = 0.1$, electron and ion density $n_e = n_i = n_0 = 5\cdot 10^{20}\,\mathrm{m}^{-3}$ with a 50–50 DT mixture, and electron temperature $T_e = 15~\mathrm{keV}$. The quasisymmetric poloidal and toroidal numbers are taken to be $M=1$ and $N=2$, respectively. A well behaved change of variables from $\vartheta, \zeta$ to $\alpha, \eta$, requires $M\neq qN$. Unless otherwise stated, we choose the poloidal and toroidal numbers of the perturbation to be $m=n=3$, so the resonance occurs around $q=m/n=1$. The presence of perturbations with different $(m,n)$ can enhance energy losses and connect their radial extent across the device. For simplicity, we take the radial location to be $r\sim r_0$, although different $r$ and $r_0$ values are considered in subsection \ref{subsection52}. In Figure \ref{figure2}, we plot the resonant plateau transport parameter $D\tau_s/r_0^2$ versus $q$, the most sensitive parameter in the resonance condition.

Panel (a) shows that transport increases near the resonant rational surface. This is due to the quantity $\left.[C_1/(\partial\lambda/\partial k)]\right|_{k_{\mathrm{res}}}$ in $D$ being maximum at $k_{\mathrm{res}}\to 0$, which happens closer to the rational surface as shown in panel (d). For each $\delta B = B_\perp^{33} = B_\parallel^{33}$, the curves are colored where the model is not restricted, gray and solid in the intervals of $q$ where the model is restricted by $\delta B_{\max}^{\mathrm{gen}}$ but not by $\delta B_{\max}^{k\to 1}$, and gray and dotted where the model is restricted by both conditions. The color dots show the values of $q$ within which the model is restricted by $\delta B_{\max}^{\mathrm{gen}}$, and the color dots with gray center the values for $\delta B_{\max}^{k\to 1}$. These intervals are set in panel (b), where we plot the maximum perturbation amplitudes $\delta B_{\max}$, in green $\delta B_{\max}^{\mathrm{gen}}$ and in red $\delta B_{\max}^{k\to 1}$. The shaded region represents the values of $q$ where $\delta B_{\max}^{\mathrm{gen}}$ restricts the model for each $\delta B$, or equivalently the values of $\delta B$ restricted for each $q$. Finally, panel (c) represents the transport obtained by using the maximum allowable perturbation from $\delta B_{\max}^{\mathrm{gen}}$ and $\delta B_{\max}^{k\to 1}$. This shows around $1\%$ energy losses when using the more restrictive condition, and around $10\%$ with the less restrictive one. For both curves, losses are well below $100\%$ so the $\delta B_{\max}^{100}$ restriction does not have an effect.

The neglected nonlinear term that limits the model is expected to smooth the peaks of the transport curves. It is reasonable to expect this effect to take place between both restrictions, but further investigation is needed. Still, the model predicts in its region of validity, for a perturbation of $\delta B = 10^{-3}~\mathrm{T}$, non negligible passing alpha energy losses from $1\%$ to $10\%$ for a range of values of $q$ that could connect ample regions of $r$. This perturbation amplitude does not seem unrealistic for building a $\bar{B}_0=10$ T stellarator, as it would require a level of precision of $\delta B/\bar{B}_0 = 10^{-4}$.

    \subsection{Sensitivity analysis}\label{subsection52}

We perform a sensitivity analysis of different parameters that affect resonant plateau transport, using $\delta B = 10^{-3}\ \mathrm{T}$.

\begin{figure}
  \centering
  \includegraphics[height=0.51\textwidth]{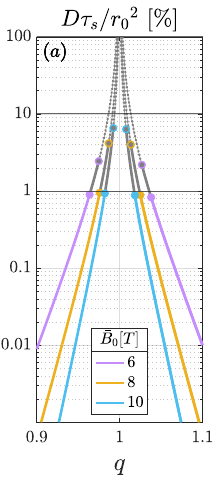}
  \includegraphics[height=0.51\textwidth]{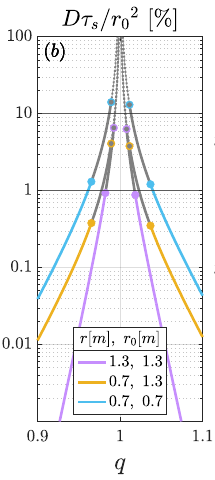}
  \includegraphics[height=0.51\textwidth]{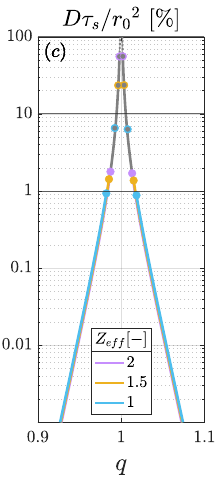}
  \includegraphics[height=0.51\textwidth]{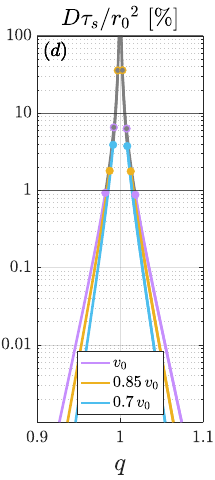}
  \caption{Sensitivity analysis of transport for $\delta B = 10^{-3}\,\mathrm{T}$. In (a) we vary $\bar{B}_0$. In (b), we change the radial location $r$ and the minor radius $r_0$. In (c), we scan different values of $Z_{\mathrm{eff}}$. In (d), we explore how losses change if $v_0$ could be decreased.}
  \label{figure3}
\end{figure}

In Figure \ref{figure3}, we explore the effect of changing the background magnetic field amplitude, the radial location and minor radius, the effective ion charge and the birth speed of the alpha particles. Panel (a) shows that lower values of $\bar{B}_0$ increase the transport curves by enhancing $\delta B/\bar{B}_0$, and also by making $\omega_{d0}$ larger, which makes the resonance easier to satisfy in wider regions of $q$ around the rational surface. At the same time, making the resonance easier to satisfy enlarges the region of $q$ where $\partial f_1/\partial r$ is relevant and the model is constrained by $\delta B_{\max}$. In the next panels, we use $\bar{B}_0 = 10\,\mathrm{T}$ unless otherwise stated. In panel (b), we decrease $r$ and $r_0$, causing a similar increase in $\omega_{d0}$ and the effect of $\delta B_{\max}$, but without increasing $\delta B/\bar{B}_0$. If we only vary $r$, this causes a decrease of transport, but if we decrease $r_0$ then $D\tau_s/r_0^2$ will grow. Panel (c) shows that higher effective atomic numbers $Z_{\mathrm{eff}}$ decrease the restriction of the model, by increasing $\nu_{\mathrm{pas}}$ and $(\Delta\lambda)_\nu$. Finally, in panel (d) we see that decreasing the alpha particle birth speed $v_0$ has a very small effect on the $D\tau_s/r_0^2$ curves, with $D$ integrated from $0$ to $v_0$. However, the constraint $\delta B_{\max}$ is calculated at $v_0$ and is very sensitive to this value, showing it to be most restrictive at higher speeds and indicating transport could be higher for particle sources at lower speeds. 

\begin{figure}
    \centering
    \includegraphics[height=0.51\textwidth]{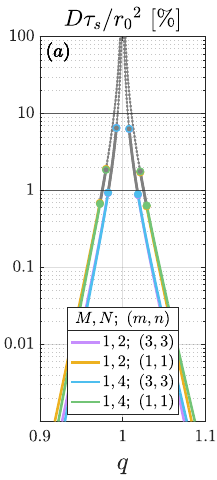}
    \includegraphics[height=0.51\textwidth]{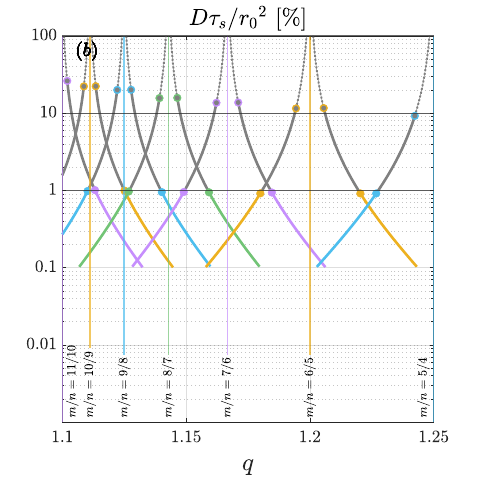}
    \includegraphics[height=0.51\textwidth]{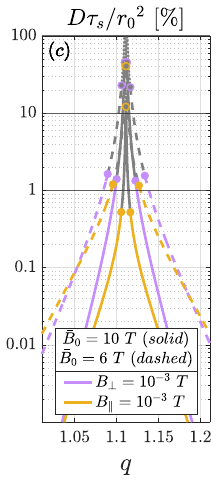}
    \caption{Sensitivity analysis of transport for $\delta B = 10^{-3}\,\mathrm{T}$. In (a), we vary the poloidal and toroidal numbers of the QS background $M$ and $N$, and of the perturbation $m$ and $n$. In (b), we present the overlapping transport from different perturbation harmonics. In (c), we show the effect of the $B_\perp$ and $B_\parallel$ pieces of a $(m,n)=(10,9)$ error field for different $\bar{B}_0$ values.}
  \label{figure4}
\end{figure}

In Figure \ref{figure4}, we explore the effect of changing the angular dependence of the background and perturbation magnetic fields, of the different parallel and perpendicular error components, and of having overlapping transport from several error fields. In panel (a), we see that lower $p^{nm}\approx n$ values increase transport but show a more restricted region of validity. In addition, for each $(m,n)$ values, the toroidal QS background number does not substantially affect transport, resulting nearly identical curves for $N=2$ and $N=4$. Panel (b) shows how transport from several error fields, plotted in different colors to help identify curves to $(m,n)$ values, can add up and overlap non negligible losses over wide regions of $q$, and consequently of $r$. In general, one mode does not affect the $\delta B_{\max}$ constraint of a different one, since for a given radial location, each mode produces an $f_1$ that lives around a different pitch angle. Since interactions between different modes are neglected, we can linearly add their transport as $\mathcal{Q}=\sum_{n,m} \mathcal{Q}_{nm}$, $\Gamma=\sum_{n,m}\Gamma_{nm}$ and $D=\sum_{n,m}D_{nm}$. Finally, in (c) we explore independently the effect of $B_\perp$ and $B_\parallel$ with $(m,n)=(10,9)$. In general, for passing alphas  and the same error field amplitude, $B_\perp$ causes much higher transport than $B_\parallel$. However, for high $p^{nm}\approx 9$ and low $\bar{B}_0$, transport from $B_\parallel$ can become comparable with that of $B_\perp$.

\begin{figure}
    \centering
    \includegraphics[height=0.51\textwidth]{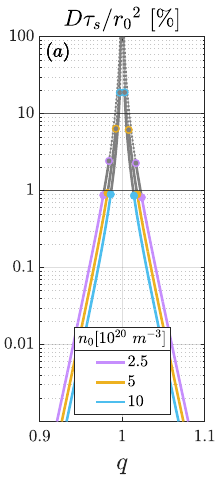}
    \includegraphics[height=0.51\textwidth]{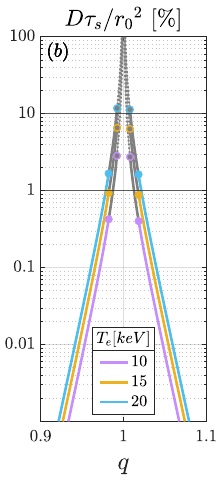}
    \includegraphics[height=0.51\textwidth]{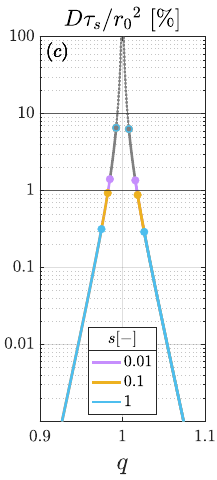}
    \includegraphics[height=0.51\textwidth]{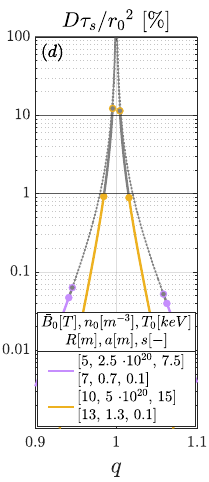}
    \caption{Sensitivity analysis of transport for $\delta B = 10^{-3}\,\mathrm{T}$. In (a) and (b) we change the density and temperature, respectively. In (c), we vary the shear. In (d), we compare the transport in a bigger stellarator similar to our baseline case, and a smaller stellarator.}
  \label{figure5}
\end{figure}

In Figure \ref{figure5}, we explore the effect of changing the density, temperature and shear, and the characteristic parameters of the stellarator. Panel (a) shows that higher density reduces $D\tau_s/r_0^{2}$ by making $\tau_s$ smaller, but it makes the $\delta B_{\max}^{k\to 1}$ condition less restrictive, which can allow higher levels of transport. In (b) we see that higher temperatures lead to more transport by making $\tau_s$ larger. In panel (c) we vary $s$, which does not change transport values since as previously noted shear effects are averaged out along the QS trajectory of passing particles. However, lower $s$ values decrease the rate of change of the resonance condition over $r$, which makes $\partial f_1/\partial r$ smaller and relaxes the $\delta B_{\max}^{\mathrm{gen}}$, which increases the transport allowed by this condition, while keeping $\delta B_{\max}^{k\to 1}$ unchanged. At the same time, the region where the model is restricted, although narrower in $q$, can be wider in $r$. Finally, panel (d) compares our baseline case with higher density, with a smaller scale stellarator. For the smaller case, we obtain slightly larger losses, but the changes in $\bar{B}_0$ and $r_0$ enhance substantially the linearization restriction, while compromising the approximation in \eqref{3.14}.
\[
\\[8pt]
\]

\section{Discussion}\label{discussion}

It has been found that non negligible passing alpha energy losses can be driven by a resonance between alpha particle streaming and drift in the vicinity of rational surfaces. By neglecting the nonlinear term in the $f_1$ equation, we perform a quasilinear evaluation of the energy flux. Our model predicts within its region of validity collisional energy losses of up to $10\%$ of the alpha energy before alpha particles have slowed down, for departures from quasisymmetry of the magnetic field direction and amplitude of $\delta B \sim 10^{-3}\,\mathrm{T}$. However, linearization is shown to lead to an incomplete description in the immediate vicinity of rational surfaces where passing resonances occur.

The most important magnetic field perturbation for the passing alpha particles is the deflection of its direction, $B_{\perp}$, rather than the change in its magnitude, $B_{\parallel}$. We considered a background field amplitude of $\bar{B}_0 = 10\,\mathrm{T}$. It is therefore realistic to assume that in real devices we can allow errors in the magnetic field of order $\delta B / \bar{B}_0 \sim 10^{-4}$.

Resonant plateau transport happens in the presence of field perturbations with poloidal and toroidal numbers $m$ and $n$, in the vicinity of rational surfaces with safety factor $q = m/n$. The magnitude of this transport depends on several factors. Some sensitive design parameters are magnetic field strength, minor radius, effective atomic number, density and temperature. As shown in Figure \ref{figure3}, lower background magnetic fields increase transport but make the linearization assumption more restrictive, smaller minor radii increase $D \tau_s / r_0^2$, and higher $Z_{\mathrm{eff}}$ values increase the transport levels by relaxing the linearization constraint. Figure \ref{figure5} shows that higher temperatures result in larger transport, and higher densities can allow enhanced transport by making the linearization constraint less restrictive.

Another sensitive design parameter is magnetic shear, as seen in Figure \ref{figure5}. If we have low shear and a resonant perturbation in the plasma, higher levels of transport are allowed by the linearization constraint, and the extent of radial locations with values of $q$ close to $m/n$ will increase. This effect can allow transport to connect larger regions of $r$ in the plasma, although the region where our treatment is limited might get wider in minor radius. Interestingly, if we have low shear and a rational surface near the edge of the plasma, as happens when using island divertors, even if the rational $q$ is not in the plasma, there will be transport associated with it since $q \approx m/n$ in the plasma. On the other hand, if we have high shear, we tend to intercept more rational surfaces resonant with error fields, but their associated transport is more localized in radius. However, if there are many error field modes their associated transport can overlap and drive losses from the core to the edge, as indicated in Figure \ref{figure4}. Then, for high shear it is best to avoid transport overlap from several error fields. In general, it is advisable to minimize $(m,n)$ perturbations around $q = m/n$ rational surfaces to decrease this type of transport.

Other observations include relaxed $\delta B_{\max}$ constraints allowing larger transport levels for higher toroidal numbers $p^{nm} \approx n$. Also, unlike trapped particles, passing alphas are particularly sensitive to errors in the magnetic field direction $B_{\perp}$, and less affected by deviations from quasisymmetry of the magnetic field amplitude $B_{\parallel}$. In addition, errors close to the edge can become a concern as it is easier for them to produce energy losses that reach the first wall, especially since optimized configuration errors tend to increase away from the magnetic axis. 

Finally, the limitations of the model show that the linearized treatment required for the quasilinear evaluation of the energy flux is harder to justify in the immediate vicinity of resonant rational surfaces. In these regions of Figures \ref{figure2}--\ref{figure5}, we have plotted our transport parameter in gray lines to indicate that the nonlinear piece neglected in the linearized drift kinetic $f_1$ equation could become relevant, using solid lines when the model is limited just by the less restrictive constraint $\delta B_{\max}^{\mathrm{gen}}$ and dashed lines when also limited by the less restrictive constraint $\delta B_{\max}^{k \to 1}$. Although it is yet to be seen in what way this term enters, it is expected to do so in a smooth way, possibly flattening the gray transport peaks (Catto \citeyear{catto_collisional_2025}), perhaps to somewhere between the values given by both constrains. Further research is required to determine the energy losses in this regime. Even though this term may decrease or saturate energy losses, the region where the linearized equation holds still predicts non negligible transport levels from passing alpha particles. In particular, high $Z_{\mathrm{eff}}$, small minor radius $r_0$ and low shear lead to larger maximum transport values.

\section*{Acknowledgements}\label{acknowledgements}

The authors acknowledge support from the Mauricio and Carlota Botton Foundation fellowship, the DOE grant DE-FG02-91-ER54109, and the Rafael del Pino Foundation scholarship.

\[\\[10pt]\]

\appendix\label{Appendix}

\section{Drifts components and canonical helical momentum}\label{appendixA}

First, we evaluate the radial component of the magnetic drift in 
$\psi_h$, $\eta$, $\alpha$ variables for a general stellarator configuration 
with well defined flux surfaces. Using \eqref{2.1}, \eqref{2.12} and 
$\nabla \times\bm{B}\cdot\nabla \psi_h = 0$, we first obtain
\[
\bm{v}_d\cdot\nabla \psi_p
= \frac{v_{\parallel}}{\Omega}\,\nabla \times
\left(\frac{v_{\parallel}}{B}\bm{B}\right)\cdot\nabla \psi_p
= \frac{v_{\parallel}}{\Omega}\,
\nabla \!\left(\frac{v_{\parallel}}{B}\right)\times\bm{B}
\cdot\nabla \psi_p
\]
\begin{equation}
= \frac{v_{\parallel}}{\Omega}
\frac{\partial}{\partial\eta}\left(\frac{v_{\parallel}}{B}\right)
\nabla \eta\times\bm{B}\cdot\nabla \psi_p
+ \frac{v_{\parallel}}{\Omega}
\frac{\partial}{\partial\alpha}\left(\frac{v_{\parallel}}{B}\right)
\nabla \alpha\times\bm{B}\cdot\nabla \psi_p .
\label{A1}
\end{equation}
We calculate 
$\nabla \eta\times\bm{B}\cdot\nabla \psi_p 
= I_h \bm{B}\cdot\nabla \eta/(M-qN)$
and 
$\nabla \alpha\times\bm{B}\cdot\nabla \psi_p = -B^2$
using \eqref{2.1}, \eqref{2.6} and \eqref{2.7}. The radial drift then becomes
\begin{equation}
\bm{v}_d\cdot\nabla \psi_p
= v_{\parallel}\,\bm{b}\cdot\nabla 
\left[\frac{I_h v_{\parallel}}{(M-qN)\Omega}\right]
- v_{\parallel} B
\frac{\partial}{\partial\alpha}\left(\frac{v_{\parallel}}{\Omega}\right) .
\label{A2}
\end{equation}
The first term comes from the quasisymmetric $\eta$ dependence of the
background magnetic field, and the second accounts for the $\alpha$
dependence due to a departure from quasisymmetry of the magnetic field
magnitude. In the drive term of the $f_0$ and $f_1$ equations, we focus
on the effect of the second term. The first term is not considered as it
causes the standard neoclassical response in a perfectly QS device that
has been studied before (Landreman $\&$ Catto \citeyear{landreman_effects_2010}) and is insensitive
to small departures from QS.

The $\nabla \eta$ component of the drift is calculated using the last
approximate form of \eqref{2.12} as
\[
\bm{v}_d\cdot\nabla \eta
= \frac{v_{\parallel}}{\Omega}
\nabla \cdot\left(
\frac{v_{\parallel}}{B}\bm{B}\times\nabla \eta\right)
= \frac{v_{\parallel}}{\Omega}
\nabla \cdot\left[
\frac{v_{\parallel}}{B}
(K_p \nabla \psi_p + G\nabla \vartheta + I\nabla \zeta)
\times\nabla (M\vartheta - N\zeta)\right]
\]
\[\\[-14pt]\]
\[
= \frac{v_{\parallel}}{\Omega}\left[
\nabla \!\left(\frac{K_p v_{\parallel}}{B}\right)\times\nabla \psi_p
\cdot\nabla \eta
- \nabla \!\left(\frac{G v_{\parallel}}{B}\right)
\times\nabla \vartheta\cdot\nabla (N\zeta)
- \nabla \!\left(\frac{I v_{\parallel}}{B}\right)
\times\nabla \zeta\cdot\nabla (M\vartheta)\right]
\]
\[\\[-14pt]\]
\begin{equation}
= v_{\parallel}\left[
\frac{\partial}{\partial\alpha}\left(\frac{K_p v_{\parallel}}{\Omega}\right)
- \frac{1}{M-qN}
\frac{\partial}{\partial\psi_p}\left(\frac{I_h v_{\parallel}}{\Omega}\right)
\right] \bm{b}\cdot\nabla \eta .
\label{A3}
\end{equation}

Using $\nabla \psi_h = (M-qN)\nabla \psi_p$, the preceding
results allow us to check the conservation of the canonical helical momentum $\psi_*$. We start by calculating
\[
(\,v_{\parallel}\bm{b}+\bm{v}_d\,)\cdot\nabla \psi_*
= (v_{\parallel}\bm{b}+\bm{v}_d)\cdot
\nabla \!\left(\psi_h - \frac{I_h v_{\parallel}}{\Omega}\right)
= - v_{\parallel}\bm{b}\cdot\nabla 
\left(\frac{I_h v_{\parallel}}{\Omega}\right)
+ \bm{v}_d\cdot\nabla \psi_h
- \bm{v}_d\cdot\nabla \left(\frac{I_h v_{\parallel}}{\Omega}\right)
\]
\[\\[-14pt]\]
\begin{equation}
= -(M-qN) v_{\parallel} B
\frac{\partial}{\partial\alpha}\left(\frac{v_{\parallel}}{\Omega}\right)
- \bm{v}_d\cdot\nabla \left(\frac{I_h v_{\parallel}}{\Omega}\right).
\label{A4}
\end{equation}
Next, we note
\begin{equation}
\bm{v}_d\cdot\nabla \left(\frac{I_h v_{\parallel}}{\Omega}\right)
= \bm{v}_d\cdot\nabla \psi_p
\frac{\partial}{\partial\psi_p}\left(\frac{I_h v_{\parallel}}{\Omega}\right)
+ \bm{v}_d\cdot\nabla \eta
\frac{\partial}{\partial\eta}\left(\frac{I_h v_{\parallel}}{\Omega}\right)
+ \bm{v}_d\cdot\nabla \alpha
\frac{\partial}{\partial\alpha}\left(\frac{I_h v_{\parallel}}{\Omega}\right)
\label{A5}
\end{equation}
\[\\[-14pt]\]
\[
= - v_{\parallel} B 
\frac{\partial}{\partial\alpha}\left(\frac{v_{\parallel}}{\Omega}\right)
\frac{\partial}{\partial\psi_p}\left(\frac{I_h v_{\parallel}}{\Omega}\right)
+ \frac{v_{\parallel}}{B}
\frac{\partial}{\partial\alpha}\left(\frac{K_p v_{\parallel}}{\Omega}\right)
\bm{B}\cdot\nabla \left(\frac{I_h v_{\parallel}}{\Omega}\right)
+ \bm{v}_d\cdot\nabla \alpha
\frac{\partial}{\partial\alpha}\left(\frac{I_h v_{\parallel}}{\Omega}\right).
\]
As a result, we get
\[
(\,v_{\parallel}\bm{b}+\bm{v}_d\,)\cdot\nabla \psi_*
= -(M-qN)v_{\parallel}B
\frac{\partial}{\partial\alpha}\left(\frac{v_{\parallel}}{\Omega}\right)
+ v_{\parallel} B
\frac{\partial}{\partial\alpha}\left(\frac{v_{\parallel}}{\Omega}\right)
\frac{\partial}{\partial\psi_p}\left(\frac{I_h v_{\parallel}}{\Omega}\right)
\]
\[\\[-14pt]\]
\begin{equation}
=- \frac{v_{\parallel}}{B}
\frac{\partial}{\partial\alpha}\left(\frac{K_p v_{\parallel}}{\Omega}\right)
\bm{B}\cdot\nabla \left(\frac{I_h v_{\parallel}}{\Omega}\right)
- \bm{v}_d\cdot\nabla \alpha
\frac{\partial}{\partial\alpha}\left(\frac{I_h v_{\parallel}}{\Omega}\right),
\label{A6}
\end{equation}
which vanishes for a quasisymmetric magnetic field.

By using the approximated expression for the drift
$\bm{v}_d\simeq\Omega^{-1}v_{\parallel}\nabla \times(v_{\parallel}\bm{b})$,
we have replaced the parallel velocity correction 
$u_{\parallel}\bm{b}
= -\Omega^{-1}v_{\perp}^2 \bm{b}\,\bm{b}\cdot\nabla \times\bm{b}$
by 
$-\Omega^{-1}v_{\parallel}^2\bm{b}\,\bm{b}\cdot\nabla \times\bm{b}$
and treated both as negligible. To verify this is reasonable, we use
\eqref{2.6} and \eqref{2.7} to obtain
\[\\[-20pt]\]
\[
B^2 \bm{b}\cdot\nabla \times\bm{b}
= \nabla \times\bm{B}\cdot
(I\nabla \zeta + G\nabla \vartheta)
= \nabla \cdot\!\left[
K_p \nabla \psi_p\times
(I\nabla \zeta + G\nabla \vartheta)\right]
\]
\begin{equation}
= \nabla \cdot\left[
\frac{K_p I_h}{M\!-\!qN}\nabla \psi_p\!\times\!\nabla \alpha
\!+\! \frac{K_p(qI\!+\!G)}{M\!-\!qN}
\nabla \psi_p\!\times\!\nabla \eta\right]
= \left[
\frac{qI\!+\!G}{M\!-\!qN}\frac{\partial K_p}{\partial\alpha}
\!-\! \frac{I_h}{M\!-\!qN}\frac{\partial K_p}{\partial\eta}
\right]\! \bm{B}\cdot\nabla \eta .
\label{A7}
\end{equation}
Neglecting $\alpha$ derivatives and noting $K_p$ is a small component of the magnetic field we find it is reasonable to ignore this correction since
\begin{equation}
\frac{\Omega^{-1}v_{\parallel}^2
(\bm{b}\cdot\nabla \times\bm{b})
(\bm{b}\cdot\nabla \psi_*)}
{\bm{v}_d\cdot\nabla \psi_h}
\sim \frac{v_{\parallel}}{\Omega}\bm{b}\cdot\nabla \times\bm{b}
\lesssim \frac{v}{\Omega R} \ll 1 ,
\label{A8}
\end{equation}
and only small terms depending on $K_p$ would be affected.

Finally, we need the tangential drift. We evaluate this drift along the full quasisymmetric trajectory using
$\alpha_*=\zeta - q(\psi_*)\vartheta = \zeta - q_*\vartheta$ 
to avoid secular behavior, and first obtain
\[\\[-22pt]\]
\[
\omega_{\alpha}
= (v_{\parallel}\bm{b}+\bm{v}_d)\cdot\nabla \alpha_*
= (v_{\parallel}\bm{b}+\bm{v}_d)\cdot\nabla \zeta
- q(\psi_*)(v_{\parallel}\bm{b}+\bm{v}_d)\cdot\nabla \vartheta
\]
\begin{equation}
= \bm{v}_d\cdot\nabla \zeta
- q_* \bm{v}_d\cdot\nabla \vartheta
- (q_*-q)v_{\parallel}\bm{b}\cdot\nabla \vartheta .
\label{A9}
\end{equation}
where the secular behaviour from 
$\mathbf{v}_d\!\cdot\nabla\alpha_* = \mathbf{v}_d\!\cdot\nabla\zeta - q_*\,\mathbf{v}_d\!\cdot\nabla\vartheta - \vartheta\,\mathbf{v}_d\!\cdot\nabla q_*$ 
is avoided by using 
$(v_{\parallel}\mathbf{b}+\mathbf{v}_d)\!\cdot\nabla\psi_* = 0$ 
from \eqref{A6} for a quasisymmetric field.
It is convenient to define $K_h = K_p/(M-qN)$. Next, we can calculate,
similarly as in \eqref{A3},
\[
\bm{v}_d\cdot\nabla \zeta
= \frac{v_{\parallel}}{B}\,
\nabla \cdot\left(
\frac{v_{\parallel}}{\Omega}\bm{B}\times\nabla \zeta\right)
= \frac{v_{\parallel}}{B}\left[
\nabla \vartheta\times\nabla \zeta
\cdot\nabla \!\left(\frac{Gv_{\parallel}}{\Omega}\right)
+
\nabla \psi_h\times\nabla \zeta\cdot
\nabla \!\left(\frac{K_h v_{\parallel}}{\Omega}\right)
\right]
\]
\begin{equation}
= \frac{v_{\parallel}}{B}\,
\bm{B}\cdot\nabla \eta\left[
\frac{\partial}{\partial\psi_h}\left(\frac{Gv_{\parallel}}{\Omega}\right)
- \frac{\partial}{\partial\vartheta}
\left(\frac{K_h v_{\parallel}}{\Omega}\right)
\right],
\label{A10}
\end{equation}
and
\[
\bm{v}_d\cdot\nabla \vartheta
= \frac{v_{\parallel}}{B}\,
\nabla \cdot\left(
\frac{v_{\parallel}}{\Omega}\bm{B}\times\nabla \vartheta\right)
= \frac{v_{\parallel}}{B}\left[
\nabla \zeta
\times\nabla \vartheta\cdot\nabla \!\left(\frac{Iv_{\parallel}}{\Omega}\right)
+
\nabla \psi_h\times\nabla \vartheta\cdot
\nabla \!\left(\frac{K_h v_{\parallel}}{\Omega}\right)
\right]
\]
\begin{equation}
= -\frac{v_{\parallel}}{B}\,
\bm{B}\cdot\nabla \eta\left[
\frac{\partial}{\partial\psi_h}\left(\frac{Iv_{\parallel}}{\Omega}\right)
- \frac{\partial}{\partial\zeta}
\left(\frac{K_h v_{\parallel}}{\Omega}\right)\right].
\label{A11}
\end{equation}
By substituting \eqref{A10} and \eqref{A11} into \eqref{A9}, and expanding
$q_* \!\simeq\! q(\psi_h) - (I_h v_{\parallel}/\Omega)\,\partial q/\partial\psi_h$,
we get
\begin{equation}
\omega_{\alpha} \!
= \! \frac{v_{\parallel}}{B}\bm{B}\cdot\nabla \eta\!
\left[
q_*\frac{\partial}{\partial\psi_h}\!\left(\!\frac{Iv_{\parallel}}{\Omega}\!\right)
\!+\! \frac{\partial}{\partial\psi_h}\!\left(\!\frac{Gv_{\parallel}}{\Omega}\!\right)
\!+\! \frac{I_h v_{\parallel}}{(M\!-\!qN)\Omega}\frac{\partial q}{\partial\psi_h}
\right]
\!- (M \!- q_* N)v_{\parallel}\bm{b}\cdot\nabla\! 
\left(\!\frac{K_h v_{\parallel}}{\Omega}\!\right) \!,
\label{A12}
\end{equation}
where in the first term we can approximate $q_* \bm{v}_d\cdot\nabla \vartheta \approx 
q \bm{v}_d\cdot\nabla \vartheta$ under the same assumption used in
\eqref{3.14}, and the second and last terms can be ignored as small to obtain
\begin{equation}
\omega_{\alpha}
= \frac{v_{\parallel}}{B}\,\bm{B}\cdot\nabla \eta
\left[
q\frac{\partial}{\partial\psi_h}\left(\frac{Iv_{\parallel}}{\Omega}\right)
+ \frac{I_h v_{\parallel}}{(M-qN)\Omega}
\frac{\partial q}{\partial\psi_h}
\right].
\label{A13}
\end{equation}

To leading order, the Vlasov operator in a quasisymmetric magnetic field
in $\psi_*, \alpha_*, \eta$ variables is
$(v_{\parallel}\bm{b}+\bm{v}_d)\cdot\nabla f_1
= v_{\parallel}\bm{b}\cdot\nabla f_1
+ \omega_{\alpha}{\partial f_1}/{\partial\alpha_*}$,
with $\omega_\alpha$ given by \eqref{A13},
using $(v_{\parallel}\bm{b}+\bm{v}_d)\cdot\nabla \psi_* = 0$
from \eqref{A13}, and neglecting as small the drift correction 
$\bm{v}_d\cdot\nabla \eta
= - v_{\parallel}\bm{b}\cdot\nabla \eta\,
\partial(I_h v_{\parallel}/\Omega)/\partial\psi_h$
to the streaming term $v_{\parallel}\bm{b}\cdot\nabla \eta$ as in \eqref{3.14}.

\section{Resonant plateau and $\sqrt{\nu}$ transport in the $k \to 1$ limit}\label{appendixB}

In this Appendix, we put the results of the model derived in this paper into context with previous models, and discuss the impact of $\sqrt{\nu}$ transport. Previous works study the  $\sqrt{\nu}$ transport and the interplay with resonant plateau transport for trapped particles close to the trapped-passing boundary (Calvo \textit{et al.} \citeyear{calvo_effect_2017}; Catto \citeyear{catto_collisional_2019}; Catto \textit{et al.} \citeyear{catto_merging_2023}). In the barely passing limit $k\to 1$, the trajectory averaged drift term, which is typically smaller than streaming, diverges as 
$K(k) \approx \ln\!\left(4/\sqrt{1-k^2}\right) \approx -\ln\!\sqrt{1-k}$, 
which seems to make the resonance with the tangential drift easier to satisfy. However, this work shows that in the vicinity of the rational surface $q=m/n$, the passing resonance is more important at smaller $k_{\mathrm{res}}$ values, which modifies the predicted energy losses.

Considering the $k\to 1$ asymptotic limit of our passing alpha model, with $k\approx 1$ and therefore $\lambda\approx 1$ and $\partial\lambda/\partial k\approx 4\epsilon$, the $f_1$ kinetic equation \eqref{3.34} becomes
\begin{equation}
f_{nm}^p\left[\sigma(qn-m)\frac{v}{qR}
+ p^{nm}\,\frac{\omega_{d0}}{2}\ln(1\!-\!k)\right]
+
i\frac{\nu_{\mathrm{pas}}}{\pi(2\epsilon^3)^{1/2}}
\frac{\partial^2 f_{nm}^p}{\partial k^2}
=
i\sigma p^{nm}v\,
\frac{\partial f_0}{\partial\psi_p}\,
A_{nm}.
\label{B1}
\end{equation}
with 
$A_{nm}
=
A_{\parallel}^{nm}
+
\sigma\left[
{v\ln(1-k)}/{2\sqrt{2}\,\epsilon\,\pi\,\overline{\Omega}_0}
\right]
B_{\parallel}^{nm}$. The resonance factor is defined as the ratio of the streaming and drift terms multiplying factors, and can be approximated close to the rational surface as
\begin{equation}
\Pi
=
\frac{\sigma(qn-m)\,2v}{p^{nm}\omega_{d0}qR}
\approx
\frac{\sigma(q-m/n)\,2v}{\omega_{d0}qR}
>0.
\label{B2}
\end{equation}
where the resonance condition requires $\sigma(qn-m)/p^{nm}\approx \sigma(q-m/n)>0$.

Defining $\gamma = e^{\Pi}$ and $x=\gamma(1-k)$, \eqref{B1} becomes
\begin{equation}
\frac{\partial^2 f_{nm}^p}{\partial x^2}
-
i\frac{\pi\epsilon^{3/2}p^{nm}\omega_{d0}}
{\sqrt{2}\,\gamma^2\nu_{\mathrm{pas}}}\,
\ln(x)\,f_{nm}^p
=
\sigma p^{nm}v\,
\frac{\pi(2\epsilon^3)^{1/2}}{\gamma^2\nu_{\mathrm{pas}}}
\frac{\partial f_0}{\partial\psi_p}
A_{nm}.
\label{B3}
\end{equation}
with the streaming plus drift terms proportional to $ln(x)$, and the resonance happening when $x=1$. We can Taylor expand around it using $ln(x) \approx x-1$. Outside the close vicinity of the resonant rational surface, $\gamma\gg 1$, so $x=1$ requires $k\to 1$. This makes $\ln(1-k)
=
\ln(x/\gamma)
\approx -\ln\gamma
$
and
$
A_{nm}
\approx
A_{\parallel}^{nm}
-
\sigma\left[
{v\ln\gamma}/{2\sqrt{2}\epsilon\pi\overline{\Omega}_0}
\right]
B_{\parallel}^{nm}.
$

\begin{figure}
  \centering
  \includegraphics[width=0.4\textwidth]{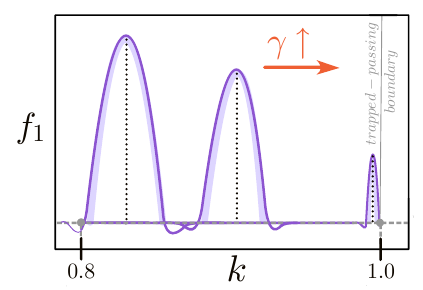}
  \caption{Perturbed distribution function as we move away from the rational surface $q=m/n$. Adapted from Catto \textit{et al.} (\citeyear{catto_merging_2023}).}
  \label{figure6}
\end{figure}

We define
$u \!=\! \left[ {\sqrt{2}\gamma^2\nu_{\mathrm{pas}}}/{\pi\epsilon^{3/2}p^{nm}\omega_{d0}} \right]^{1/3}\!\!$
, with $u$ generally small, and let $z \!=\! (x\!-\!1)/{u}$, so that \eqref{B3} becomes
\begin{equation}
\frac{\partial^2 f_{nm}^p}{\partial z^2}
- iz\,f_{nm}^p = \frac{(2\epsilon^3)^{1/2}\pi u^2 p^{nm}v}
{\gamma^2\nu_{\mathrm{pas}}}
\frac{\partial f_0}{\partial\psi_p}
A_{nm}.
\label{B4}
\end{equation}
By letting
\begin{equation}
f_{nm}^p
=
-\frac{2v}{\omega_{d0}u}
\frac{\partial f_0}{\partial\psi_p}
A_{nm}
\,\Upsilon(z),
\label{B5}
\end{equation}
the kinetic equation becomes an inhomogeneous Airy equation
\begin{equation}
\frac{\partial^2\Upsilon}{\partial z^2}
-
iz\,\Upsilon
=
-1,
\label{B6}
\end{equation}
whose particular solution is (Su $\&$ Oberman \citeyear{su_collisional_1968})
\begin{equation}
\Upsilon_{\mathrm{SO}}
=
\int_0^{\infty}
d\tau\,e^{-iz\tau - \tau^3/3}
\; \xrightarrow[|z|\gg 1]{} \;
\frac{-i}{z}
=
\frac{-iu}{x-1}
.
\label{B7}
\end{equation}
\[\\[20pt]\]
This solution vanishes away from the resonance $x=1$ as in the right side of \eqref{B7} and is $\Upsilon_{\mathrm{SO}}(z=-1/u)\approx iu$ at the trapped-passing boundary. In the main text of this paper, $\gamma$ was not defined and did not need to be asymptotically large, so $u$ would have been very small and $\Upsilon_{\mathrm{SO}}$ negligible at the trapped-passing boundary. In this Appendix, as we move away from $q=m/n$ and require $k\to 1$, $\gamma\gg 1$ becomes an asymptotically large parameter. Consequently, $u$ is generally small, but not negligible, and the particular solution does not satisfy the boundary condition $\Upsilon(z=-1/u)=0$. This boundary condition requires us to keep the homogeneous piece of the solution as in Catto \textit{et al.} (\citeyear{catto_merging_2023}), which introduces $\sqrt{\nu}$ transport. In the extreme case that $\Pi\gg 1$, $\gamma\gg 1$ becomes so large that $u\gg 1$, and the $\sqrt{\nu}$ piece might dominate (Calvo \textit{et al.} \citeyear{calvo_effect_2017}; Catto \citeyear{catto_collisional_2019}). This transition is depicted in Figure \ref{figure6}.

To calculate the homogeneous piece of the small $u$ regime, we let $z=iy=ye^{i\pi/2}$ so that \eqref{B6} becomes $\partial^2 \Upsilon/\partial y^2 - y\Upsilon = 1$. This equation has the homogeneous solution $\Upsilon_{\sqrt{\nu}} = C\,\mathrm{Ai}(y e^{i2\pi/3} = z e^{i\pi/6})$ (Abramowitz $\&$ Stegun \citeyear{abramowitz_handbook_1964}), with $\mathrm{Ai}(y e^{i2\pi/3})$ an Airy function vanishing more rapidly than $\Upsilon_{\mathrm{SO}}$ as we move away from the resonance and $z\gg 1$. By imposing the boundary condition $\Upsilon(z=-1/u)=0$ in the full solution $\Upsilon=\Upsilon_{\mathrm{SO}}+\Upsilon_{\sqrt{\nu}}$, we calculate $C = -{\Upsilon_{\mathrm{SO}}(z=-1/u)}/{\mathrm{Ai}(y e^{i2\pi/3} = -e^{i\pi/6}/u)},$ with $\mathrm{Ai}(y e^{i2\pi/3} = -e^{i\pi/6}/u)\approx (u^{1/4}/\pi^{1/2}) e^{i\pi/24} \sin\!\left[\, 2(1+i)/3u^{3/2} + \pi/4 \right]$. The full solution is (Tolman $\&$ Catto \citeyear{tolman_drift_2021})
\begin{equation}
\Upsilon = \Upsilon_{\mathrm{SO}}(z) + \Upsilon_{\sqrt{\nu}}(z)
= \Upsilon_{\mathrm{SO}}(z) - \Upsilon_{\mathrm{SO}}(-1/u)\,
\frac{\mathrm{Ai}(z e^{i\pi/6})}{\mathrm{Ai}(-e^{i\pi/6}/u)} .
\label{B8}
\end{equation}
We evaluate the transport caused by this solution, where $a_{\parallel}(k\to 1) = (qn-m)/2\pi q$, as
\begin{equation}
D_{u<1} \!
= \! \sum_{n,m} \!\frac{\pi\epsilon v_0^2}{p^{nm}\,\ln(v_0/v_c)\,{\omega}_0}
\!\left[
\frac{e^{-\Pi_0}}{2+\Pi_0}
\!+\! \frac{1}{3}\!
\left( \!
\frac{{a'}/{\rho_0}}{\pi q p^{nm}}
\frac{\nu_0 R}{v_c}
\! \right)^{1/2}
\right] \!
\left[
\! \left( \! \frac{B_{\perp}^{nm}}{\bar{B}_0} \! \right)^2
\!+ \!
\bar{q}^2
\left( \! \frac{B_{\parallel}^{nm}}{\bar{B}_0} \! \right)^2
 \right] \! , \! \! \! 
\label{B9}
\end{equation}
with $\bar{q} = (qn-m) p^{nm}$, $\Pi_0 = \Pi(v=v_0)$ and $\nu_0 = v_{\lambda}^3/\epsilon\tau_s v_0^3$.

\begin{figure}
  \centering
  \includegraphics[width=0.74\textwidth]{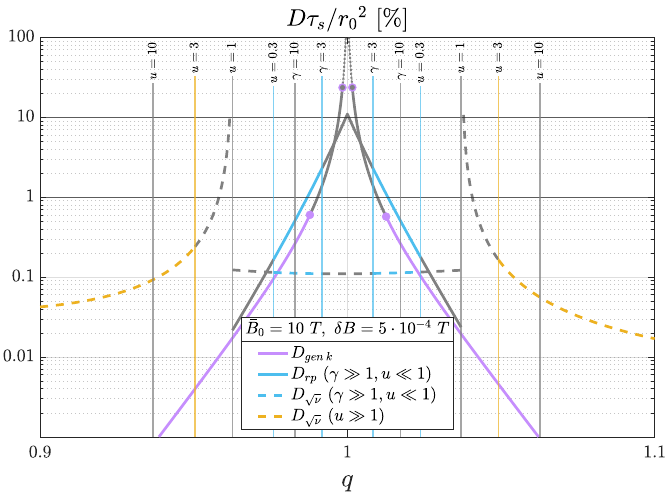}
  \\[-8pt]
  \caption{Comparison of $\sqrt{\nu}$ and resonant plateau transport for the different models in this paper.}
  \label{figure7}
\end{figure}

In the case $u\gg 1$, equation \eqref{B3} becomes $\partial^2 F/\partial x^2 - i u^{-3} \ln(x) F = -1$. An adequate solution for $x\sim u^{3/2}\gg 1$ is (Calvo \textit{et al.} \citeyear{calvo_effect_2017}; Catto \citeyear{catto_collisional_2019})
\begin{equation}
f_{nm}^p = -\frac{2v}{\omega_{d0} u^3}
\frac{\partial f_0}{\partial \psi_p} A_{nm}\, F
\approx
\frac{2v}{\omega_{d0}}
\frac{\partial f_0}{\partial \psi_p} A_{nm}\,
\frac{i}{\ln(u^{3/2})}
\left(
1 - e^{-(1+i)x\sqrt{\ln(x)/(2u^3)}}
\right),
\label{B10}
\end{equation}
which gives an energy diffusivity, with $u_0=u(v=v_0)$,
\begin{equation}
D_{u>1} \!
= \! \sum_{n,m}
\frac{2\epsilon v_0^2}{3[\ln(u_0^3)]^{3/2}\,\ln(v_0/v_c)\,{\omega}_0}
\!\left[
\frac{2\pi ({a'}/{\rho_0})}{q (p^{nm})^{3}}
\frac{\nu_0 R}{v_c}
\right]^{1/2} \!
 \!\left[ \!
\left(\frac{B_{\perp}^{nm}}{\bar{B}_0}\right)^2
+
\bar{q}^2
\left(\frac{B_{\parallel}^{nm}}{\bar{B}_0}\right)^2
\right].
\label{B11}
\end{equation}
\[\\[-24pt]\]
In Figure \ref{figure7}, we compare the model in the main text and the asymptotic solutions in this Appendix, using the parameters of our baseline case stellarator and a perturbation amplitude $\delta B = 5\cdot 10^{-4}\ \mathrm{T}$. The curves are colored within their regime of validity and gray outside. The purple curve represents the resonant plateau transport for the general $k_{\mathrm{res}}$ model presented in this paper, noted here as $D_{\mathrm{gen}\,k}$. As explained in Figure \ref{figure2}, this curve is solid gray when limited by $\delta B_{\max}^{\mathrm{gen}}$ and dashed gray when also limited by $\delta B_{\max}^{k\to 1}$. The blue curves in Figure \ref{figure7} show the case $\gamma\gg 1\gg u$, with the resonant plateau and $\sqrt{\nu}$ pieces of $D_{u<1}$ plotted with solid and dashed lines, respectively. Away from the rational surface, the $k_{\mathrm{res}}$ calculated for $D_{\mathrm{gen}\,k}$ gets closer to~1, making the asymptotic expansion $k\to 1$ in $D_{\mathrm{rp}}$ more precise and showing similar resonant plateau transport in both models. Closer to the rational surface, $k_{\mathrm{res}}$ is reduced, and the general case $D_{\mathrm{gen}\,k}$ can predict larger transport very close to $q=m/n$. Moreover, in the regime of validity $\gamma\gg 1\gg u$, we see that the $\sqrt{\nu}$ transport is smaller than the resonant plateau. It is reasonable to extend this observation to the general case solved in the main text, since $k_{\mathrm{res}}$ does not need to be close to~1, so the homogeneous piece is not needed to make $f_1$ vanish at the trapped--passing boundary and the $\sqrt{\nu}$ transport associated with it is negligible. Lastly, in orange we represent the $\sqrt{\nu}$ transport $D_{u>1}$ from the $\gamma\gg u\gg 1$ case. Within its range of validity, $D_{u>1}$ gives transport up to values similar to the $\sqrt{\nu}$ piece of $D_{u<1}$ closer to the rational surface, and decays away from the rational surface. In general, passing transport decreases when we get away from the rational surface, as $f_1$ is suppressed by being pushed towards the trapped--passing boundary, where it has to vanish. We conclude that the model from the main text is more general and precise, and that resonant plateau transport dominates over $\sqrt{\nu}$ transport.

\[\\[-12pt]\]

\section{Pitch-angle integral}\label{appendixC}

In the main text we solve the following integral in velocity space
\[
\int_0^{1-\epsilon} d\lambda \int_0^\infty d\tau\, e^{-\tau^3/3} \cos(Z\tau)
=\int_0^1 dk\, \left.\frac{\partial\lambda}{\partial k}\right|_{\epsilon}
\int_0^\infty d\tau\, e^{-\tau^3/3} \cos[\Xi(k-k_{\mathrm{res}})\tau]
\]
\begin{equation}
\approx \left.\frac{\partial\lambda}{\partial k}\right|_{\epsilon, k=k_{\mathrm{res}}}
\int_0^1 dk \int_0^\infty d\tau\, e^{-\tau^3/3} \cos[\Xi(k-k_{\mathrm{res}})\tau]
= \frac{\pi}{\Xi} \left.\frac{\partial\lambda}{\partial k}\right|_{\epsilon, k=k_{\mathrm{res}}} ,
\label{C1}
\end{equation}
where $Z=\Xi(k_{\mathrm{res}}-k)$, with $\Xi\gg 1$, so $k=k_{\mathrm{res}}-\Xi^{-1} Z$. To prove \eqref{C1}, we can Taylor expand $\left.{\partial\lambda}/{\partial k}\right|_{\epsilon}$, since it varies slowly over $k$ compared to $\cos[\Xi(k-k_{\mathrm{res}})\tau]$, as
\begin{equation}
\left.\frac{\partial\lambda}{\partial k}\right|_{\epsilon,\,k=k_{\mathrm{res}}-Z/\Xi}
=\left.\frac{\partial\lambda}{\partial k}\right|_{\epsilon,\,k=k_{\mathrm{res}}}
-\frac{Z}{\Xi} \left.\frac{\partial}{\partial k}\left(\frac{\partial\lambda}{\partial k}\right)\right|_{\epsilon,\,k=k_{\mathrm{res}}}
+\cdots .
\label{C2}
\end{equation}
The integral associated with the first term would be
\[
I_1 = \int_0^1 dk \int_0^\infty d\tau\, e^{-\tau^3/3} \cos[\Xi(k-k_{\mathrm{res}})\tau]
= \frac{1}{\Xi} \int_0^\infty d\tau\, e^{-\tau^3/3}
\int_{-(1-k_{\mathrm{res}})\Xi}^{k_{\mathrm{res}}\Xi} dZ\, \cos(Z\tau) 
\]
\begin{equation}
\! \! = \frac{1}{\Xi} \! \! \int_0^\infty \! \! \! d\tau\, e^{-\tau^3/3}\,  \!\frac{1}{\tau} 
\{ \sin(\Xi k_{\mathrm{res}}\tau) + \sin[\Xi(1-k_{\mathrm{res}})\tau] \}
= \frac{1}{\Xi} \! \! \int_0^\infty \! \! \! d\tau\, e^{-\tau^3/3}\, 2\pi\delta(\tau)
= \frac{\pi}{\Xi},
\label{C3}
\end{equation}
where $k_{\mathrm{res}}\Xi \gg 1$, $(1-k_{\mathrm{res}})\Xi \gg 1$. The integral associated with the second term goes as
\[I_2 = \int_0^1 dk\, \frac{Z}{\Xi}
\int_0^\infty d\tau\, e^{-\tau^3/3} \cos[\Xi(k-k_{\mathrm{res}})\tau]
= \frac{1}{\Xi^2} Re \!\left[
\int_{-(1-k_{\mathrm{res}})\Xi}^{k_{\mathrm{res}}\Xi} dZ\,
Z \int_0^\infty d\tau\, e^{-\tau^3/3 - iZ\tau} \right]
\]
\[
= \! \frac{1}{\Xi^2} Re \!\left[
i \int_{-(1-k_{\mathrm{res}})\Xi}^{k_{\mathrm{res}}\Xi} \!\!\!\!\!\!\! dZ
\int_0^\infty \!\!\!\!\! d\tau\, e^{-\tau^3/3} \frac{d}{d\tau} e^{-iZ\tau} \right]   
\!=\! - \frac{1}{\Xi^2} Re \!\left[
\int_0^\infty \!\!\!\!\!d\tau\, \tau e^{-\tau^3/3}
\! \! \int_{-(1-k_{\mathrm{res}})\Xi}^{k_{\mathrm{res}}\Xi}
\!\!\!\!\!\!\! dZ\, \frac{d}{dZ} e^{-iZ\tau} \right]
\]
\begin{equation}
= - \frac{1}{\Xi^2}
\int_0^\infty d\tau\, \tau e^{-\tau^3/3}
\{ \cos(\Xi k_{\mathrm{res}}\tau)
- \cos[\Xi(1-k_{\mathrm{res}})\tau] \}.
\label{C4}
\end{equation}
We can bound this integral by letting $x = \Xi K \tau$ where $K=k_{\mathrm{res}}$ or $1-k_{\mathrm{res}}\sim \mathcal{O}(1)$ or smaller
\[
\int_0^\infty d\tau\, \tau e^{-\tau^3/3} \cos(\Xi\tau K)
= \frac{1}{(\Xi K)^2}
\int_0^\infty dx\, x\, e^{-(x/\Xi K)^3/3} \cos(x)
\approx \frac{1}{(\Xi K)^2}
\int_0^{K\Xi} dx\, x\cos(x)
\]
\begin{equation}
= \frac{1}{(\Xi K)^2}
[\cos(\Xi K) + (\Xi K)\sin(\Xi K)]
\sim \mathcal{O}(1/\Xi K) \sim \mathcal{O}(1/\Xi).
\label{C5}
\end{equation}
Therefore $I_2 \sim \mathcal{O}(1/\Xi^3)$, and we can conclude that
\[
\int_0^{1-\epsilon} d\lambda \int_0^\infty d\tau\, e^{-\tau^3/3} \cos(Z\tau)
= \left.\frac{\partial\lambda}{\partial k}\right|_{\epsilon,\,k=k_{\mathrm{res}}} I_1
- \left.\frac{\partial}{\partial k}\left(\frac{\partial\lambda}{\partial k}\right)\right|_{\epsilon,\,k=k_{\mathrm{res}}} I_2 + \cdots
\]
\begin{equation}
= \frac{\pi}{\Xi} \left.\frac{\partial\lambda}{\partial k}\right|_{\epsilon,\,k=k_{\mathrm{res}}}
+ \mathcal{O}(1/\Xi^3)
\approx \frac{\pi}{\Xi} \left.\frac{\partial\lambda}{\partial k}\right|_{\epsilon,\,k=k_{\mathrm{res}}},
\label{C6}
\end{equation}
which has also been checked numerically. This would be equivalent for $f(k)=\left.{(\partial\lambda}/{\partial k)}\right|_{\epsilon}/C_1$, and $f(k)$ could come as a prefactor outside of the integral, evaluated at $f(k_{\mathrm{res}})$.

\section{The model validity constraints}\label{appendixD}

As discussed in section \ref{section4}, the linearization constraint $V<\tilde{\omega}_d (\Delta\lambda)_\nu (\Delta r)_\nu$ allows us to evaluate when the nonlinear term of the $f_1$ equation can become relevant. This condition sets a maximum perturbation amplitude $\delta B_{\max}$ that depends on the radial position and gets more restrictive in the nearest vicinity of the resonant rational surface. Typically, two behaviors dominate the evolution of this condition, which are depicted on the left panel of Figure \ref{figure1}. Closer to the rational surface, the rapid variation of the resonant pitch angle will dominate, which will result in a fast radial dependence in $f_1$ and a small $\Delta r$. Farther away from the rational surface, $k_{\mathrm{res}}(\lambda_{\mathrm{res}},\epsilon)$ will be approximately $1$, which will fix the relation between the adiabatic invariant $\lambda_{\mathrm{res}}$ and the radial location $\epsilon$, allowing us to obtain the $f_1$ radial scale $\Delta r$ from its pitch angle width $\Delta\lambda$.

To calculate $\Delta r$ we use \eqref{4.9} and \eqref{4.10} to obtain
\begin{equation}
\Delta r
=
\frac{\displaystyle \frac{\omega_0}{q^2}
\frac{\left.{\partial F}/{\partial k}\right|_{\epsilon}}{
\left.{\partial \lambda}/{\partial k}\right|_{\epsilon}}
\Delta\lambda_{\mathrm{res}}}
{\displaystyle
\frac{v_0}{qR} \frac{\partial q}{\partial r}
\frac{\partial}{\partial q}
\left(
\left| qn-m \right|
\frac{M-qN}{q^2}
\right)
-
\frac{\omega_0}{q^2} \frac{1}{R}
\left(
\left.\frac{\partial F}{\partial k}\right|_{\epsilon}
\left.\frac{\partial k}{\partial \epsilon}\right|_{\lambda}
+
\left.\frac{\partial F}{\partial \epsilon}\right|_{k}
-\frac{3}{2}\frac{F}{\epsilon}
\right)
}
.
\label{D1}
\end{equation}
By using $C_1=\left.\partial F/\partial k\right|_{\epsilon}$ from \eqref{3.39}, $\left.\partial\lambda/\partial k\right|_{\epsilon}$ as defined after \eqref{3.32}, and calculating
\begin{equation}
\frac{\partial}{\partial q}
\left(
\left| qn-m \right|
\frac{M-qN}{q^2}
\right)
=
n \frac{M-qN}{q^2}
-
\left| qn-m \right|
\frac{(2M-qN)}{q^3} ,
\label{D2}
\end{equation}
\begin{equation}
\left.{\partial k}/{\partial \epsilon}\right|_{\lambda}
=
{k(2-k^2)}/{4\epsilon} ,
\label{D3}
\end{equation}
\begin{equation}
\left.\frac{\partial F}{\partial \epsilon}\right|_{k}
=
-
\frac{(2-k^2)(3k^2-2\epsilon k^2+4\epsilon)}
{2\big[(1-\epsilon)k^2+2\epsilon\big]\big[(1-2\epsilon)k^2+4\epsilon\big]}
F ,
\label{D4}
\end{equation}
\[\\[4pt]\]
we can calculate the maximum perturbation amplitude in \eqref{4.15} as
\begin{equation}
\delta B_{\max}^{\mathrm{gen}}
=
\left|
\frac{\bar{B}_0}{v_0 \big(1\!+\!a_{\parallel}/a_{\perp}\big)}
\frac{
\displaystyle
\frac{R^2 \epsilon}{v_0 n s}\,
\tilde{\omega}_d^{\,2} (\Delta\lambda)_\nu^{\,2}
}{
\displaystyle
\left(\!
1\!-\!\frac{|qn\!-\!m|}{qn}\frac{2M\!-\!qN}{M\!-\!qN}
\!\right)
\!-\!
\frac{R\epsilon}{v_0 n s}\,
p^{nm}\omega_0 \!
\left(
C_1 \frac{k(2\!-\!k^2)}{4\epsilon}
\!+\!
\frac{p(k)}{2\epsilon} G
\right)
}
\right| .
\label{D5}
\end{equation}
with
$p(k)
=
\big[{
k^4(3-6\epsilon+4\epsilon^2)
+
k^2(12\epsilon-16\epsilon^2)
+
16\epsilon^2
}\big]/{
k^2 \big[(1-\epsilon)k^2+2\epsilon\big]^{3/2}
}$
and $s=(r/q)\,\partial q/\partial r$. The collisional boundary layer width has been taken to be $(\Delta\lambda)_\nu\approx 2(\tilde{\nu}_{\mathrm{pas}}/\tilde{\omega}_d)^{1/3}$, from the estimates in section \ref{section4} and looking at the precise shape of $\Upsilon_{\mathrm{SO}}$.

As mentioned at the end of section \ref{section4}, the behavior for $k_{\mathrm{res}}\to 0$ is somewhat uncertain, since the approximation in \eqref{3.33} for the collision operator is compromised in this limit. Besides, as seen in Figure \ref{figure2}, $k_{\mathrm{res}}\to 1$ happens not too far from the rational surface. Consequently, another sensible constraint is
\begin{equation}
\delta B_{\max}^{k\to 1}
\equiv
\delta B_{\max}^{\mathrm{gen}}(k_{\mathrm{res}}\to 1)
=
\left|
{R\tilde{\omega}_d (\Delta\lambda)_\nu^{\,2} \bar{B}_0}/
{v_0 \big(1+a_{\parallel}/a_{\perp}\big)}
\right| .
\label{D6}
\end{equation}

For reference, we can also calculate the behavior in the limit $k_{\mathrm{res}}\to 0$, which as discussed before, exhibits the variation in $k_{\mathrm{res}}$ when approaching the rational surface
\begin{equation}
\delta B_{\max}^{\mathrm{var}}
\equiv
\delta B_{\max}^{\mathrm{gen}}(k_{\mathrm{res}}\to 0)
=
\left|
{R^2 \epsilon \tilde{\omega}_d^{\,2} (\Delta\lambda)_\nu^{\,2} \bar{B}_0}/
{v_0^2 n s \big(1+a_{\parallel}/a_{\perp}\big)}
\right| .
\label{D7}
\end{equation}

We can take a more extreme limit by making $k_{\mathrm{res}}\approx 0$. Assuming this limit in $\tilde{\nu}_{\mathrm{pas}}$, $a_{\parallel}$ and the asymptotic expansion of $\tilde{\omega}_d$, then \eqref{D7} becomes
\begin{equation}
\delta B_{\max}^{k\to 0}
\equiv
\delta B_{\max}^{\mathrm{gen}}(k_{\mathrm{res}}\approx 0)
=
\left|
\frac{4R^2 \epsilon \bar{B}_0}
{v_0^2 n s \big(1+v_0/\bar{\Omega}_0 r_0\big)}
\left(
\frac{\pi}{2}
\sqrt{\nu_{\mathrm{pas}}}\,
p^{nm} \omega_0 \epsilon k_{\mathrm{res}}
\right)^{4/3}
\right| .
\label{D8}
\end{equation}

\[\\[14pt]\]

\begin{figure}
  \centering
  \includegraphics[width=0.5\textwidth]{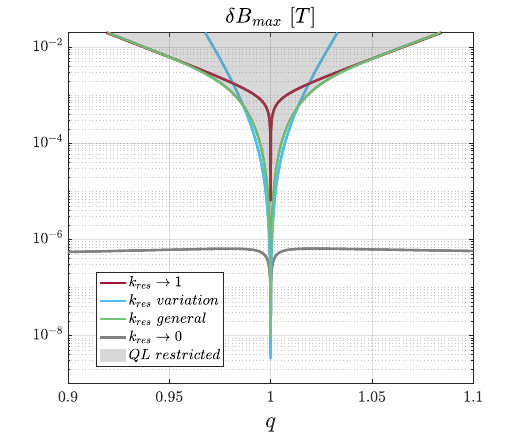}
  \caption{Maximum perturbation amplitude constraints: $\delta B_{\max}^{k\to 1}$ (red), $\delta B_{\max}^{\mathrm{var}}$ (blue), $\delta B_{\max}^{\mathrm{gen}}$ (green) and $\delta B_{\max}^{k\to 0}$ (gray).}
  \label{figure8}
\end{figure}

In Figure \ref{figure8}, we plot the different perturbation amplitude constraints \eqref{D5}-\eqref{D8}. The more general calculation of the constraint is $\delta B_{\max}^{\mathrm{gen}}$. For every location $q$, this linearization constraint limits perturbation amplitudes above $\delta B_{\max}^{\mathrm{gen}}$, which we shade in gray. However, since the collision operator approximation requires us to be slightly away from $k_{\mathrm{res}}\to 0$, and $k_{\mathrm{res}}$ goes to $1$ very fast as one gets away from the rational surface, another sensible constraint is $\delta B_{\max}^{k\to 1}$. As expected, $\delta B_{\max}^{\mathrm{gen}}$ converges asymptotically to $\delta B_{\max}^{k\to 1}$ away from the rational surface. As can be seen in $\delta B_{\max}^{\mathrm{var}}$, the difference between $\delta B_{\max}^{k\to 1}$ and $\delta B_{\max}^{\mathrm{gen}}$ comes precisely from the variation of $k_{\mathrm{res}}$ as one gets closer to the rational surface and $k_{\mathrm{res}}$ decreases towards $0$. Finally, the limit $k_{\mathrm{res}}\approx 0$ in $\delta B_{\max}^{k\to 0}$ indicates the behavior and minimum perturbation amplitude almost right at the rational surface, where a minimum value of $k_{\mathrm{res}}\approx10^{-2}$ has been considered.

As shown in Figures \ref{figure2}-\ref{figure5} and \ref{figure7}, the transport allowed by the model is very sensitive to the choice of $\delta B_{\max}$. Since both $\delta B_{\max}^{\mathrm{gen}}$ and $\delta B_{\max}^{k\to 1}$ constraints seem sensible, we check both in the Figures. It is reasonable to assume that the nonlinear term will saturate the transport levels (Catto \citeyear{catto_collisional_2025}), presumably somewhere in between the values given by these two constraints.
\[\\[-20pt]\]

\bibliographystyle{jpp}
\bibliography{Paper}

\end{document}